\documentstyle[a4wide,11pt]{article}
\input mssymb
\newcommand{\be}{\begin{equation}}
\newcommand{\ee}{\end{equation}}
\newcommand{\bea}{\begin{eqnarray}}
\newcommand{\eea}{\end{eqnarray}}

\newcommand{\raw}{\quad \longrightarrow \quad}
\newcommand{\impl}{\quad \Longrightarrow \quad}

\newcommand{\ld}{\quad \ldots \quad}
\newcommand{\nd}{\noindent}
\newcommand{\beqa}{\be \begin{array}{rcl}}
\newcommand{\eeqa}{\end{array} \ee}

\newcommand{\tld}{\widetilde}
\newcommand{\no}{\nonumber}
\newcommand{\gf}{{{\Bigl\arrowvert}_{{\lbrace{t_{{\tld  N},i}}\rbrace}
 \,  = \,  0}}}

\newcommand{\grd}{\nabla}

\newcommand{\dfn}{\quad  \stackrel{\rm def}{=}   \quad }
\newcommand{\vek}{\overrightarrow}
\newcommand{\ben}{\begin{enumerate}}
\newcommand{\een}{\end{enumerate}}
\newcommand{\bi}{\begin{itemize}}
\newcommand{\ei}{\end{itemize}}
\newcommand{\bd}{\begin{description}}
\newcommand{\ed}{\end{description}}

\catcode `\@ = 11

\title{{\bf {$2d$ Gravity  Coupled to Topological Minimal Models}}}

\bigskip

\bigskip
\bigskip

\bigskip

\author{{\sf Avijit Mukherjee} \thanks{{\it E-mail}:
a.mukherjee@damtp.cambridge.ac.uk}\\
\\
{\it Department of Applied Mathematics and Theoretical Physics} \\
{\it University of Cambridge},\\
{\it Silver Street},\\
{\it Cambridge \,   CB3    9EW, \,  U.K.}\\}

\date{{\it Revised},  \, November    1994}

\begin{document}
\maketitle

\begin{abstract}

We discuss the  properties of genus-0 correlation functions of a
topological {\it minimal} model (the model ${A_{k + 1}}$) coupled
 to $2d$ topological gravity. The action of the {\it twisted}
minimal model is  perturbed by  non-trivial couplings to gravitational
 descendants,  which,  in turn,  are constructed entirely from the
 fields in the   matter sector of the theory. We develop an explicit
 formulation, in terms of orthogonal polynomials, for investigating
 the {\it large phase space}  of the theory. Some useful identities for
 correlation functions (valid on the {\it large phase space}) of the
 theory are  established   and the {\it puncture} and {\it dilaton equations}
  of topological gravity are obtained as special cases of these general
 relations. Finally, we obtain  an important relation expressing the
 gravitational couplings in terms of the couplings in the {\it small
 phase space}, ({\it i.e.},  the couplings to the chiral primaries).
 Thus  eventually we are able to solve for the coordinates (couplings)
 of the large phase space  in terms of the LG superpotential
 characterizing the matter sector of the model.

\end{abstract}

\vspace{-15.5cm}

\begin{flushright}
DAMTP-94/30\\
\end{flushright}
\vspace{15.5cm}

\newpage

\section{Introduction}

Topological field theories {\cite{WIT1}  are a class of exactly
solvable systems {\it without}  local propagating degrees of freedom.
  These are, even {\it  without}  coupling to gravity,  generally
{\it covariant} quantum field theories.  It would, however, be
incorrect to  think that these theories are then of merely mathematical
 interest,  without having any real physical relevance. In fact,
 in the last few years,  we have realized that these theories not
 only do possess interesting  physical properties, but are also
  related at a  more fundamental level to other useful physical
theories like the matrix models and $d = 1$ string theory.

One of the handy prescriptions for constructing  a large class of
useful  topological conformal field theories (TCFT) is to start with
a $N= 2$ superconformal field theory (SCFT) and then  {\it twist}
(or {\it antitwist} it)  it according to some definite set of rules
 {\cite{EY}}, {\cite{DVV1}} to obtain a TCFT. In particular, if one
 initially starts with a $0 \, <  \, c  \, <  \, 3$ {\it minimal}
$N= 2$ model,   then one  has   a TCFT  with a finite dimensional
physical Hilbert space. It is known that  a $N = 2$ SCFT has a special
 set  of states-- the so called {\it chiral primary states}, which
 form a multiplicative  chiral ring.  Further when these theories
also have a  convenient Landau-Ginzburg (LG) description,  the chiral
 ring then becomes the finite-dimensional  polynomial  (in the LG fields)
 ring  modulo the equation of motion {\cite{{EM},{VW},{LVW}}}.
 Under the twisting procedure, it is this chiral ring which becomes
 identified with the Hilbert space of physical states of the topological
 theory-- {\it i.e.} all the other states in the original $N = 2$ model
 {\it decouple}  from the Hilbert space of the topological theory.
 However, after {\it coupling}  these theories to gravity,  this
 picture  changes  completely   and the Hilbert space of physical
 states becomes  {\it infinite-dimensional}.

For topological field theories constructed from $N = 2$ superconformal
 field theories, the underlying chiral ring structure of the latter
 translates into an {\it associative} operator algebra for the 3-point
 correlation function of the topological theory. The 3-point
correlation function plays a very fundamental role  in topological
 field theory as all other correlation functions  can be determined
 in terms of it. Thus this associativity constraint  provides us,
 in principle, with sufficient information to obtain all the correlation
 functions of the theory. If further, the topological theory be
 perturbed by elements of the chiral ring, the operator algebra
 still retains the essential structure (even though the chiral
ring is then no longer nilpotent). The structure  constants of
the multiplicative chiral ring algebra then  become functions
of the perturbing parameters (the couplings) -- and their
associativity property (which luckily is still retained
even in the perturbed theory) once  again gives us sufficient
constraints to determine the 3-point  (and hence in general $N$-point)
 correlation functions of the perturbed topological model.
However from the practical computational point of view, the
above approach does not provide a useful prescription at all.
It is at this point that we find that the use of orthogonal
polynomials most facilitating {\cite{{DVV1}}}. We therefore
 adopt this approach throughout in our work. We shall try to
generalize this method to the case when we couple our model to
topological gravity.  It turns out that  in this case  too, the
 orthogonal polynomials prove to be of invaluable help in
investigating the correlation functions  and other properties of
 the gravity-coupled model.

Topological field theories possesss a nilpotent symmetry $Q$,
and the stress-tensor is $Q$-exact. All novel properties
 {\cite{{WIT1},{DVV1}}} of these theories stem from this
characteristic feature. The physical states are the $Q$-cohomology
classes.   Further, in this work  we have considered as our model
of the matter theory, the twisted minimal $A_{k + 1}$ model.
This has a very convenient representation in terms of a single
Landau-Ginzburg field $x$, and hence the physical Hilbert space
will be a finite-dimensional space spanned by the polynomials in $x$.
  Topological gravity is the topological theory associated with the
moduli space of Riemann surfaces.   In $2d$ topological gravity,
one considers a theory with the $2d$ metric $g_{{\mu}{\nu}}$ as the
dynamical  variable, and a classical action given by a vanishing
Lagrangian:
$${{\cal L}_{classical}}  \quad  =  \quad 0$$
The above Lagrangian has more symmetries that the usual
diffeomorphism invariance. Gauging these symmetries (by
introducing ghost fields and ghosts for ghosts)-- then gives us
the model of {\it topological gravity}.  We shall  consider
coupling  such a model to topological matter theories.  The coupling
to topological gravity requires the covariantization of the
stress-tensor and the super stress-tensor of the matter sector
and also extending  our $Q$ symmetry by adding up the contribution
from the gravity sector. The resultant $Q_{total}  \, =  \,
Q_{top. matter}  \, +  \, Q_{top. gravity}$  then constitutes
our nilpotent BRST charge of our $gravity + matter$ topological
 theory. The physical Hilbert space, which is now the cohomology
class of the much enlarged $Q_{total}$ charge -- then becomes
infinite dimensional. The  $Q_{total}$-cohomology classes will
 be generically called the  {\it gravitational descendants}.
 From BRST analysis   of the physical spectra {\cite{{EK},{KL}}},
  we know that the chiral primary fields of the matter sector
 remain physical observables even after  coupling to gravity.
 They however become {\it gravitationally dressed} as   we shall
be seeing in subsequent sections (from eq.({\ref{propc}})).  When
 the topological minimal models are coupled to topological gravity
 {\cite{KL}}, some of the BRST exact states in the matter
sector\footnote{By {\it  matter sector}, we shall refer to the
twisted Hilbert space of our $N = 2$ models and this consistes
 of the (left)-chiral primary states only.}    {\it no longer
decouple}  from  the physical Hilbert space. These states now
 describe what we shall call the   {\it gravitational descendants},
 and will now constitute  the much-enlarged (infinite-dimensional)
 phase space of observables of our theory
{\cite{{DVV2},{VV},{DW},{WIT2}}}.

{}From the works of {\it Losev et. al.}  the {\it remarkable} fact
 that emerges  is that all gravitational descendants can be
constructed {\it entirely}  from fields in the matter sector
 {\cite{{LOS},{KL}}}.  Further, following {\it Eguchi et. al.},
 we construct these descendant fields by using orthogonal polynomials.
  We then enlarge the phase space of our minimal theory  to an
infinite-dimensional one by  coupling the minimal action to all
 such (infinite) gravitational descendants by introducing a coupling
for each of these fields. Thus in effect, we treat the minimal
topological  action in an infinite-dimensional  {\it background} of
   gravitational descendants, {\it  i.e.} the gravitational
descendant
(secondaries)  sector acts as  {\it external sources}  in our
formulation.  Using  our construction for the   gravitational
descendants,  we  can obtain useful Ward-identities in the
large phase space. It is known that an interesting consequence
of the use of orthogonal polynomials in topological field theory
 (without gravity), is that they  enable  us to {\it determine}
the exact forms of the couplings characterizing the strengths
of the perturbations to the minimal action {\cite{DVV1}}.
One of the main conclusions of our present work is  to
establish that the above property   can be extended even
when gravity {\it is present}. This means that in our general
model of {\it topological matter + topological gravity},  one
 can still determine,  (eventually in terms of the
superpotential characterizing the $N = 2$  matter sector) all
the couplings to the background fields (consisting of the
chiral primaries and the   gravitational descendants).
Thus the superpotential emerges as possessing  {\it all}
the wealth of informations needed to  completely investigate
such theories. This remarkable  fact lends evidence to the
assertion  that  $2d$ gravity is an  {\it induced}  effect.
 Further, if we consider the limit $k  \, {\rightarrow} \,
0$ of our results, the matter sector becomes trivial  and
we are then left with a pure gravity theory.
Thus in the {\it limit} when the dynamics of the purely
matter sector become {\it trivial}, we have,  from our work,
 a {\it topological  solution} to the problem of {\it pure}
$2d$  gravity.

The paper is organized as follows. In Sect. 2  we give  a
brief summary of mostly known results for the case of the
perturbed $A_{k + 1}$ model (without gravity), and as such
this section is essentially a brief recapitulation  of results
 from the existing literature on perturbed  Landau-Ginzburg TCFT.
  Then in Sect. 3,  we couple our model to gravity  and
introduce our constructions for the gravitational descendants.
Sect. 4 contains some useful properties for correlation
functions and other identities that can be readily derived
from our reduction formula obtained in Sect. 3. We also
relate  and compare  our results to the gravity-free case by
setting  all the gravitational couplings to zero at the end of
the day. In the same  section, we  also obtain   explicit
expressions  for the 2 and 3-point functions for our general
gravity-coupled and perturbed model. In Sect. 5, we give some
detailed analysis for recursion relations for the correlation
 functions  valid throughout  the {\it large phase space}. We
 then  obtain the generalized puncture  and dilaton equations
as a special case of these relations. In Sect. 6,  we show how
we can explicitly determine  all  the  gravitational couplings
of the theory in terms of the superpotential characterizing the
 matter sector. Since the couplings to the chiral primaries  are
 already known, this means that effectively we now know all the
 couplings in terms of the perturbed superpotential. Thus all
information about the {\it gravity-sector} also  stems  from
the superpotential characterizing the {\it matter sector}.
Hence in a sense,  this establishes that topological $2d$
gravity  is an   {\it induced}  effect  (and the case of
{\it pure}  gravity simply corresponding to  the
$k \, {\rightarrow} \,  0$ limit of the matter sector).
  Finally, in Sect. 7   we summarize our conclusions
and comment on the possible future developments.
In the appendices, we clarify some of the  calculational
 details used in our work.

\bigskip

\bigskip


\section{Solution of the Perturbed  $A_{k + 1}$ Model
Using Orthogonal Polynomials --- A Quick Recapitulation}

\setcounter{equation}{0}

Let us consider the minimal model  of the type  $A_{k + 1}$
corresponding to the {\it A}- series of the {\it ADE}
classification. This is  defined by a single Landau-Ginzburg field
$x$ characterized by the superpotential $W_0$, and  the {\it central
 charge}  $c$  given by:
\be
{W_0}({x})   \quad    =     \quad  {{x^{k + 2}}\over{k + 2}},
 \quad \quad \quad c \quad     =     \quad  {3k\over{k + 2}}
\ee
The $(k + 1)$  chiral primaries   which become the physical
operators in the corresponding {\it twisted} version
 {\cite{{EY},{DVV1}}} of the topological theory, are of the
form: $x^i  =  {\lbrace}{1, x, {x^2}, \ldots  ,{x^{k}}}{\rbrace}$.

In the unperturbed theory, the polynomials  ${{\phi}_i}({x_a}) \,
 {\in}  \,  {\cal R}$  coincide exactly with the basis of the
chiral ring\footnote{And the chiral primaries are identified
 with ${\phi_i} = {x^i}, \quad  (i = 0,1,2...k)$ and generate
the ring \,   ${\cal R}$:
\[{{{\phi}_i}{{\phi}_j}} =  \left  \{  \begin{array}{ll}
                         {\phi_{i + j}}   &   \quad
                             \quad {i + j   \leq  k}\\
                         0                &   \quad
                                    \quad {i + j > k}
                         \end{array}
                         \right.  \]} and so the
2-point function, on genus-0,  can be readily computed
using the prescription due to {\it Vafa}  {\cite {VA}} as:
\bea
{{\eta}_{ij}^{(0)}}  \quad   =  \quad
<{{\phi}_i}(x_a){{\phi}_j}(x_a)>  &   =    &
 {\oint  {dx}} \,  {{{{\phi}_i}(x_a){{\phi}_j}(x_a)}\over{{\partial_x}W}}
 \quad      =   \quad   \oint {dx}  \,
 {{{x^i}{x^j}}\over{x^{k + 1}}}\nonumber\\
 &  =  &  {{\delta}_{i + j , k}}
\eea
Hence on the sphere the operator ${\phi}_i$
 is conjugate to ${\phi}_{k - i}$. The 2-point
function (in genus-0) defines (the superscript 0
is used to remind us  that presently we are dealing
 with the gravity free case) the {\it flat} topological
 metric on the space of the primaries.

\subsection{The Perturbed Model  }

For each of the operators ${\phi}_i  = {\lbrace}{1, x, {x^2},
 {x^3},  \ldots  ,{x^{k}}}{\rbrace}$, we now  introduce
   couplings $\{t_i\}  = {\lbrace}t_0, t_1,t_2, \ldots
,t_k{\rbrace}$ with $U(1)$ charges (their canonical scaling
dimension):  ${q_{t_i}}  \,  =  \, {\left[1 -  {i\over{k + 2}}\right]}$,
 and we  consider perturbing the model as:
\bea
W(x_a ;{\underline{t}}) \quad  &  = &    \quad
 {{x^{k + 2}}\over{k + 2}} \quad  - \quad  {\sum_{j = 0}^{k}} \,
 {t_j}{x^j}\no\\
{{\phi}_i}(x_a)      \quad  &  =    &   \quad  -
 {{\partial}\over{\partial{t_i}}}  W(x_a ;{\underline{t}})
 \quad =  \quad  {x^i}\no
\eea
where ${x^a} =  {\lbrace}{1, x, {x^2}, \ldots ,{x^{k}}}{\rbrace}$,
 and the notation $\underline{t}$ is  presently used
as a short form to denote the set $\lbrace{t_i}\rbrace$
 defined above. However this is not the most general form
of the perturbation that we can consider,
 and we demand that  our perturbed theory needs to be
invariant under $t$-{\it  reparametrizations}. Thus if
we allow $t$-dependent basis transformations,  (such
redefinitions like:  ${t_i} {\raw} {g_i}({\underline t})$,
 correspond simply to reparametrizations of the couplings)--
 the metric  ${\eta_{ij}}$ will obviously lose its nice
property  of $t$-{\it independence}.
However, luckily  the space of topological field theories
is blessed with {\it zero} curvature\footnote{ One can,
in fact, set up a covariant formulation in the space of
the perturbing coupling constants, in which the ordinary
derivatives ${{\partial}\over{\partial{t_j}}}$  are replaced
by covariant ones with  respect to the couplings. It then
transpires that the Christoffel symbols in these covariant
 derivatives are the necessary {\it contact }  terms, that
 arise in constructing the theory with a general choice of
operators.} and hence there {\it does}  exist a {\it preferred}
 (modulo constant $t_j$-translations)  parametrization of
the couplings-- for which the 2-point function is  {\it constant}
   ({\it i.e.}  the metric ${\eta^{0}_{ij}} \, =  \,
 <{\phi_i}{\phi_j}>$ is {\it constant},  when considered
as a function of the moduli  $t_i$). These are the so-called
{\it flat-coordinates} on the  space of couplings $\{ {t_i} \}$.
  We shall therefore always {\cite{BV}}  make such a choice
  of coordinates $t_i$ in our work. The use of orthogonal
polynomials in fact provides us with an explicit construction
 prescription for such flat coordinates in the space of the
perturbed theory.

Armed with the above knowledge,  we may, more generally,
 consider perturbing the potential as:
\be
W(x_a ; {\underline{t}}) \quad  =  \quad  {{x^{k + 2}}\over{k + 2}}
 \quad  -  \quad  {\sum_{j = 0}^{k}} \, {g_j}({\underline{t}})\,{x^j}
\ee
where ${g_j}({\underline{t}})$ are {\it a priori}
arbitrary functions of ${\lbrace}t_i{\rbrace}$,  and the
polynomials ${{\phi}_i}(x_a ; \underline{t})$
\be
{{\phi}_i}(x_a ; {\underline{t}})      \quad   \stackrel{\rm def}{=}
  \quad  -  {{\partial_i}W(x_a ;{\underline{t}})},  \quad
 \quad \quad   ({\hbox{where}} \quad   {{\partial}_i}  =
 {{\partial}\over{{\partial}{t_i}}})
\ee
now generate the chiral ring {\it via} the   multiplication
 rule:
$$
{{{\phi}_i}(x_a ;  {\underline{t}})}\,{{\phi}_j}(x_a ;
 {\underline{t}})  \, =   \,
{\sum_l}\,{{{{\cal C}_{ij}}^l}({\underline{t}})}\,
{{\phi}_l}(x_a ; {\underline{t}})   \quad \quad
 \quad ( \,
 {\hbox{mod}}  \quad
{{{\partial}{W(x_a;{\underline{t}})}}\over{\partial}x_b}  \,   )
$$
with the perturbed structure constants
${{{\cal C}_{ij}}^l}({\underline t})$
  still satisfying the same associativity constraints:
$$
 {\sum_m}\,{{{\cal C}_{ij}}^m}({\underline{t}})\,
{{\cal C}_{mkl}}\,({\underline{t}})   \, =  \,
 {\sum_m}\, {{{\cal C}_{ik}}^m}({\underline{t}})\,
{\cal C}_{mjl}\,({\underline{t}})
$$

{}From the general properties of the correlation
functions, (Ward identities), we know that the
2-point function (which also defines the metric)
 remains unaltered even in the presence of perturbations
($i.e.$, is ${\underline{t}}$- independent).   Hence the
2-point function defined by ${{\phi}_i}(x_a ; {\underline{t}})$
must {\it coincide}  with that defined by ${\phi_i}(x_a)$.
Hence we have the  identification:
\bea
{\eta_{ij}^{(0)} }({\underline{t}})  \quad &  =  &
<{{{\phi}_i}(x_a ;  {\underline{t}})}\,{{\phi}_j}(x_a ;
{\underline{t}})>  \quad =   \quad     <{{{\phi}_i}(x_a  )}\,
{{\phi}_j}(x_a  )>\no\\
&   =   &   {\eta_{ij}^{(0)}} (0)   \quad  =  \quad
{\delta_{i + j , k}}
\eea

\subsection{Orthogonal Polynomials}

Let us now introduce a  {\it generating function }
${L_0}(x_a;{\underline{t}})$,    (once again, the subscript
 zero is used to remind us that currently, we are dealing
 with the gravity-free case)  defined by:
\be
W(x_a ; {\underline{t}})  \quad     \stackrel{\rm def}{=}
\quad    {{{L_{0}^{k+ 2  }}(x_a ; {\underline{t}})}\over{k + 2}}
 \quad    =      \quad     {{x^{k + 2}}\over{k + 2}}  \quad  -
\quad  {\sum_{j = 0}^{k}} \, {g_j}({\underline{t}})\,{x^j}
\ee
We can    then  define the $(k  +  1)$  {\it orthogonal polynomials}
\, ${{\Phi}_i}{(x_a ; {\underline{t}})}$ by:
\be
{{\Phi}_i}{(x_a ; {\underline{t}})}  \quad   \stackrel{\rm def}{=}
\quad   {1\over{i + 1}}\,
{{\left[{{\partial}\over{{\partial}x}}{L_{0}^{i + 1}}\right]}_+}
  \quad  \quad   \quad  \quad   \quad  \quad (i = 0, 1, 2,
\ldots  ,k.)
\ee
where the subscript $+$ indicates a truncation of the series
to only positive powers of $x$.   These polynomials are
{\it orthogonal}  with respect to the definition of the
inner product  in the ring ${\cal R}$.
One immediate consequence of the above definition is:
$$
{{\Phi}_0}{(x_a ; {\underline{t}})}  \quad = \quad {\bf 1}
$$
Then it is not difficult to establish the {\it identification}:
\be
 {{\Phi}_j}(x_a;{\underline{t}})  \quad \quad  {\equiv}
\quad \quad  {{\phi}_j}(x_a;{\underline{t}})
\ee
The fundamental  correlation function in the perturbed theory
known to be  the 3-point function
-- $< {{\phi}_i}{(x_a ; {\underline{t}})}
 {{\phi}_j}{(x_a ; {\underline{t}})}
{{\phi}_l}{(x_a ; {\underline{t}})}>$.
Setting one of the indices to zero,  then defines the
 metric of the theory, \, $
< {{\phi}_i}{(x_a ; {\underline{t}})}
{{\phi}_j}{(x_a ; {\underline{t}})} {{\Phi}_0}{(x_a ; {\underline{t}})}>
  \quad  =  \quad  {{\eta}_{ij}^{(0)}}( {\underline{t}})
$
But in the absence of gravity, $ {{\phi}_0}{(x_a ; {\underline{t}})}
 \, =  \, {\bf 1}$, and the above expresion {\it reduces}  to
the expression for the  2-point function, {\it i.e.} \,
$ < {{\phi}_i}{(x_a ; {\underline{t}})}{{\phi}_j}{(x_a ; {\underline{t}})}>
 \quad  =  \quad  {{\eta}_{ij}^{(0)}}( {\underline{t}})
$.
The {\it flat-coordinates}  (the couplings)  are then known
to be given by:
\be{t_{k - i}}  \quad  = \quad  {- \,{1\over{i + 1}}}\,
{\left({\hbox{res}}\,{L_{0}^{i + 1}}\right)}
\ee
Once again the derivation of the above  result relies on
 the {\it t-independence} properties of the 2-point function
of the theory.

\nd
{\bf Note:} \,    The analogous result for the couplings for
the chiral primaries with gravity {\it switched on} is (for
an explanation of the notations used please see the next section):
\be
\label{coupling}
{t_{0, k - i}}  \quad  = \quad  {- \,{1\over{i + 1}}}\,
{\left({\hbox{res}}\,{L^{i + 1}}\right)}   \quad  =  \quad  - \,
 { \, \, \, {{\left(k + 2\right)}^{{i + 1}\over{k + 2}}}\over {i + 1}}
 \, {\hbox{res}} \, {\left({W^{{i + 1}\over{k + 2}}}\right)}
\quad   \quad  \quad  {\hbox{for}} \, \, 0  \, \,
{\leq}  \, \, i  \, \, {\leq}  \,  \, k
\ee

\bigskip

\bigskip


\section{The Perturbed  and Gravity-Coupled Model}
\setcounter{equation}{0}

We shall now consider coupling our topological  matter
theory (corresponding to the `twisted' version of the
$A_{k + 1}$ model) to topological gravity. The primary
 fields in the {\it purely}  matter sector  will still
be the chiral primary fields ${\phi_i} \, (i = 0, 1, 2,
{\ldots}  , k)$, (and are in one-to-one correspondence with
 the elements of the perturbed chiral ring of the $N = 2$ model)
 and after coupling to gravity, the complete\footnote{It is believed,
 though it has not yet been proved rigorously that the complete  set
 of physical operators in {\it topological gravity} \, +  \,
{\it topological matter} does consist of the set  ${\lbrace} \,
 {\sigma_N}{\left({\phi_i}\right)}\, {\rbrace}$. In minimal
topological $2d$ gravity, there are operators ${{\sigma}_N}, \,
 N = 0, 1, 2, {\ldots},$ of {\it ghost number}  $(2N - 2)$ which
 correspond to  $(2N - 2)$-dimensional submanifold of moduli space.}
 spectrum of physical operators of our theory will be given by the
 above chiral primaries together with all (can be infinite in number)
their {\it gravitational descendants}   ${\lbrace} \,
{\sigma_N}{\left({\phi_i}\right)}\, {\rbrace}$.  These
  ${\lbrace} \, {\sigma_N}{\left({\phi_i}\right)}\, {\rbrace}$
are the BRST invariant operators with respect to the total BRST
charge of the gravity-coupled model.

In a free-field formulation of pure topological gravity,
the (super) conformal gauge fixed  action consists of the
following set of dynamical fields: the Liouville field $\rho$
and its superpartner $\psi$, the associated anti-ghosts $\pi$
and $\chi$, as well as the usual spin $(2, -1)$ Faddeev-Popov
$(b,c)$ ghost fields and their superpartners ${\left({\beta},
 {\gamma}\right)}$.  The  fundamental BRST invariant operator
 comes from the ghost sector and is
$$
{{\gamma}_0}  \quad  =  \quad  {1\over2} \,  {\left(
{\partial}{\gamma}  \, +  \,  {\gamma} {\partial}{\beta}
 \, - \, c{\partial}{\psi}\right)}  \, \, - \, \, c.c
$$
and the corresponding complete set of non-trivial physical
observables in the gravity-sector are the family of operators
defined by:
\be
{{\sigma}_N}  \quad = \quad {{{\gamma}_0}^N} \,  {\cdot} \,
 {\bf P}  \quad  \quad {\hbox{where}}  \quad {\bf P}  \, \,
=  \, \, c \, {\overline c} \,  {\delta} {\left({\gamma}\right)}
\,  {\delta} {\left({\overline  {\gamma}}\right)}  \quad \quad
\quad (N = 0, 1, 2, {\ldots})
\ee
and ${\bf P}$ is the {\it puncture } operator. After coupling
the above theory to a topological matter system, the complete
spectrum of (total) BRST invariant operators of the coupled
system are given by:
$$
 {\lbrace} \, {\sigma_{N,i}} \,  {\rbrace}     \quad {\equiv}
\quad       {\lbrace} \, {\sigma_N}{\left({\phi_i}\right)}\,
{\rbrace}  \quad  =  \quad  {\phi_i} \, {\cdot} \, {{{\gamma}_0}^N} \,
  {\cdot} \,  {\bf P} \quad \quad \quad (i = 0, 1, 2, {\ldots}  k)
$$
where ${\phi_i}$ are the left (or right) moving chiral primary
fields from the matter sector. These fields  ${\sigma_{N,i}}$ are
our gravitational descendants and their constrction and properties
have been discussed in details in the existing literature
{\cite{{DVV2},{VV},{LOS},{DW},{WIT2}}} and so we do not elaborate
 on these issues  here. The operators which are the gravitational
descendants of the {\it identity} operator are those that are present
in a model of pure topological gravity ({\it i.e.}, before any
coupling to matter models).

 It is known  {\cite{LOS}}  that these descendants of the chiral
primaries can be constructed {\it entirely} from fields in the
 matter sector (which  may consist  of the twisted version of
some $N  =  2$ theory, or topological sigma model). We shall
further call the special operator  ${\sigma_0}{\left({\phi_0}\right)}
\, =  \, {\phi_0}  \, =  \, {\bf P}$ the {\it puncture operator}
{\cite{WIT2}} and this operator will play a vital role\ in our
future discussions.

With hindsight, we may make the following observation here.
As  in the gravity-free case studied in the earlier section,
the fundamental correlation function in the small phase space
(defined later) is  once again the 3-point function: \,
$< {{\phi}_i}{(x_a ; {\underline{t}})} {{\phi}_j}{(x_a ; {\underline{t}})}
 {{\phi}_l}{(x_a ; {\underline{t}})}>$. Setting one of
its indices to zero then defines the {\it metric} of the theory:
\be
\label{defmet}
{\eta_{ij}}  {\dfn}  <{{\phi}_i}{(x_a ; {\underline{t}})}
 {{\phi}_j}{(x_a ; {\underline{t}})}
{{\phi}_0}{(x_a ; {\underline{t}})}> \quad  = \quad
  < {{\phi}_i}{(x_a ; {\underline{t}})}
{{\phi}_j}{(x_a ; {\underline{t}})} \,  {\bf P}  >
\ee
 However with non-vanishing coupling to the gravitational
descendants, ${\bf P} \, {\not=} \, {\bf 1}$, and hence in
this case (quite unlike the gravity-free case), this does not
 coincide with the 2-point function.  Thus  {\it pure
topological gravity } is the case  in which the  only
primary field in the theory is the {\it puncture operator}
 and in this case the complete spectrum of physical states
of the theory consists of: \{$ {\phi_0} \, =  \, {\bf P}$,
together with  all its  gravitational descendants
${\sigma_N}{\left({\bf P}\right)}$\}.

Introducing the infinite set of couplings    (the space of
{\it all} couplings ${\lbrace}{t_{N,i}}{\rbrace}$  will be
called the {\it phase space} of the theory)  ${\underline t} \,
= \,  {\lbrace} \, {t_{N,i}} \, {\rbrace}$,  coupling to the
$ {\sigma_N}{\left({\phi_i}\right)}$'s,  where $N = 0, 1, 2,
\ldots, {\infty}$, and $i = 0, 1, 2,  \ldots, k$, our generic
 action will look like:
\be
{\cal S}  \quad  = \quad {{\cal S}_0}  \, -  \, {\sum_{N,j}}\,
 {t_{N,j}} \,  {\int} {\sigma_N}{\left({\phi_j}\right)}
\ee
where  $ {{\cal S}_0}$   is the {\it minimal}  gravitational
action {\it plus} the action of the topological  $A_{k + 1}$
model, {\it i.e.} \, $ {{\cal S}_0} \, \,  =  \, \,
 {{\cal S}_{N = 2}}  \, +  \, {{\cal S}_{top. grv}}$.
We are thus studying the $A_{k + 1}$ topological model
(on the sphere) in the presence of an infinite-dimensional
background of gravitational descendant fields which we treat
as external sources   with relative strengths determined by
the couplings ${t_{N,i}}$.  The {\it finite} dimensional
 phase space with ${t_{N,i}}  \, =  \, 0, \, \, {\hbox{for}} \,
 N \, >  \, 0$ (which is the same as
${\lbrace}{t_{{\tld N},i}}{\rbrace}  \,  =  \,  0 $)
 plays a very important role in our theory and we shall
 call this the {\it small phase space}. Thus the
{\it small phase space} is a $(k + 1)$- dimensional
 phase space with affine coordinates ${t_{0,i}}$.
The  {\it small phase space}  thus describes the
moduli space of topological field theories that can
be reached by (relevant and marginal) perturbations of
the minimal models.  The operator
${\sigma_{0}}{\left({\phi_0}\right)}  \, =  \,
 {\phi_0}  \, =  \, {\bf P}$ is  the {\it puncture
 operator}, and its coupling $t_0$ plays the role
of the cosmological constant. The operator
${\sigma}_1$   is   called the {\it    dilaton operator}
  (borrowing terminology from string theory).
The puncture operator produces the crucially
important contact  terms when inserted in correlation functions.

In terms of the superpotential characterizing the
theory this translates into considering a perturbed
superpotential of the form:
\be
W{\left({x_a}; {\underline t}\right)}   \quad   \,
\, =    \, \quad   {{x^{k + 2}}\over{k + 2}}  \, -
\, {\sum_{N = 0}^{\infty}}  \, \,  {\sum_{i = 0}^k}\,
\,  {t_{N,i}} \, {\sigma_N}{\left({\phi_i}\right)}
\ee
where ${\sigma_N}{\left({\phi_i}\right)}$  is the $N^{th}$
 {\it  gravitational descendant} of the chiral primary
field ${\phi_i}{\left({x;{\underline t}}\right)}$,
\bea
{\sigma_N}{\left({\phi_i}\right)}   &  \quad    =
\quad  &   -\, {{\partial W}\over{\partial{t_{N,i}}}}\\
\label{dfpha}  {\phi_i}   &  \quad    =       \quad  &
  -\, {{\partial W}\over{\partial{t_{0,i}}}}   \quad  =
 \quad  - \,  {{\partial W}\over{\partial{t_{i}}}}
\eea

\bigskip
\nd
{\bf  Notation used  :}  \quad  Let us first try to clarify
 the {\it notation}  that we are going to use in our work.
\ben
\item Except when otherwise mentioned the subscript
indices $i, N, {\tld N}$  can run over the following
range of values:
\bea
i \quad  &  \quad   =  \quad  &   0, 1, 2, {\ldots} ,
k.\no\\
N \quad  &  \quad   =  \quad  &   0, 1, 2, {\ldots} ,
{\infty}\no\\
{\tld N} \quad  &  \quad   =  \quad  &   1, 2,  3, {\ldots} ,
  {\infty}\no
\eea

\item   The quantities ${\underline t}$  will henceforth
refer to the complete set of coupling parameters for the
 {\it perturbed}  and {\it gravity-coupled} model. Thus
\bea
{\underline t} \quad  & =   &  {\lbrace}{t_{N,i}}{\rbrace}
\quad  \quad  \quad \quad   \quad \quad  (N = 0, 1, 2,
{\ldots} , {\infty} \, ; \, i = 0, 1, 2, {\ldots} , k)\no\\
&  =  &  {\lbrace}  {t_{0}}, {t_{1}}, {t_{2}}, {\ldots} ,
{t_{k}}, {t_{1,0}}, {t_{1,1}}, {t_{1,2}}, {\ldots} ,
{t_{1,k}}, {\ldots} ,   {t_{N,0}}, {t_{N,1}},{t_{N,2}},
 {\ldots} ,{t_{N,k}} , {\ldots} {\rbrace}\no
\eea

\een

\nd
Allowing arbitrary reparametrizations in the coupling
constants space (the {\it large} phase space)
${\lbrace}t_{N,i}{\rbrace}, \, N = 0, 1, 2, 3,  {\ldots},$
and $i = 0, 1, 2, {\ldots}, k$, \,  the most general form
of the perturbed and gravity-coupled superpotential is:
\be
W{\left({x_a}; {\underline t}\right)}   \quad     \,  \, =
   \,    \quad   {{x^{k + 2}}\over{k + 2}}  \, -  \,
{\sum_{N = 0}^{\infty}}  \, \,  {\sum_{i = 0}^k}\, \,
 {g_{N,i}}{\left({\underline t}\right)} \, {\sigma_N}{\left({\phi_i}\right)}
\ee
where   ${g_{N,i}}{\left({\underline t}\right)}$ are once
again a priori arbitrary functions of the couplings ${\underline t}$.
The {\it generating function}  $L{\left({x_a}, {\underline t}\right)}$
 is now defined by:
\be
W{\left({x_a}; {\underline t}\right)}       \quad
 \stackrel{\rm def}{=}   \quad
 {{{L^{k + 2}}{\left({x_a}; {\underline t}\right)}}\over{k + 2}}
   \quad  =  \quad     {{x^{k + 2}}\over{k + 2}}  \, \,  -  \,
 \, {\sum_{N = 0}^{\infty}}  \, \,  {\sum_{i = 0}^k}\, \,
 {g_{N,i}}{\left({\underline t}\right)} \,
 {\sigma_N}{\left({\phi_i}\right)}
\ee

\nd
Then the $N^{th}$  gravitational  descendant \,
 $ {\sigma_N}{\left({\phi_i}\right)}$  of the
 chiral primary field ${\phi_i}$ is given by:
\be
\label{dfop}
{\sigma_N}{\left({\phi_i}\right)}     \quad
\stackrel{\rm def}{=}        \quad
{{b{\left(N, i\right)}}\over{ a(N) +  i  +  1}} \,
  {{\left[{\partial_x} L^{ a(N) +  i  +  1}\right]}_+}
\ee
Further, we also have the identification: \,
 $  {\sigma_0}{\left({\phi_i}\right)} \, =  \,
 {\phi_i},  \, \quad  {\forall}  \, \, i  \,
\quad   0 \,\,  {\, \leq} \,\,  i \,\,  {\leq} \,\,  k$.
  From consistency requirements:
\quad ${\sigma_N}{\left({\phi_i}\right)}  \,  \,  {\equiv}  \,
   \, 0, \quad {\hbox{for}}  \, \, N  \, <  \, 0, \, \,
 {\hbox{and}}  \, \,  \, {\forall}  \, \,  \, i.$ and we
   further demand  that the as yet arbitrary numerical
 functions $a(N)$ and $b(N,i)$  also satisfy the
constraints\footnote{The  above constraints may be
 readily solved to give:
\bea
a(N)  &  \quad =  \quad     &  N{\left(k + 2\right)}\no\\
 b(N,i)  &  =  &  {{\left(k + 2\right)}^{-N}} \,
{\left[ {{{\Gamma}(\lambda)}\over{{\Gamma}{\left({\lambda}
+ N\right)}}}\right]}\\
&  \quad  &  \quad \quad \quad    \quad  \quad
 {\left(N = 0,1,2, {\ldots}  {\infty} ;  \,    \,
 {\hbox{and}} \,  \,  {\lambda}  \, =  \,  {\left({{i + 1}\over{k + 2}}\right)}
\right)}\no
\eea} :
\bea
a(N)  \, -  \, a(N - 1)   &  \quad   =  \quad  &
 {\left(k + 2\right)}\nonumber\\
 {\left[ \, {b(N,i)\over{a(N)  +  i  +  1}} \, \right]}
 &  \quad   =  \quad  &  b(N + 1,i)
\eea

\bigskip

\bd
\item[Note:]   Using the definitions ({\ref{dfpha}})
 and ({\ref{dfop}}) we can readily  obtain the equation:
$$
- \, {{\partial W}\over{\partial {t_i}}}  \quad =
\quad  { \,   {{\left(k + 2\right)}^{ {{i + 1}\over{k + 2}}}} \over{{\left(i +
1\right)}}} \,{{\left[ {\partial_x}  {W^{{i + 1}\over{k + 2}}}\right]}_{+}}
$$
So that {\it relabelling} the couplings ${t_i}  {\raw}
 {\tau_i}$, where
$$
{\tau_{j - 1}}  \quad   {\dfn}  \quad  - \,
 {{\left(N\right)}^{-{j\over{N}}}} \,  j  \, {t_{j - 1}}
\quad \quad  \quad \quad   ({\hbox{where}} \quad  N =  k + 2)
$$
we can rewrite the above equation in the form:
\be
{{\partial}\over{\partial{\tau_p}}} \, W  \quad =
 \quad  {{\left[{\partial_x}{W^{{p/N} }}\right]}_{+}}
  \quad \quad  ( { \hbox{for}} \quad   p = 1,2,...,N - 1.)
\ee
which is essentially similar to the equation for the
$p$-th  primary KdV flows.

\ed
\bigskip

\nd
The correlation function
$<{\sigma_N}{\left({\phi_i}\right)}{\phi_j}>$
can be evaluated  by using the prescription  {\cite{VA}}
due to {\it Vafa} and we eventually get a useful
reduction formula  due to {\it Eguchi et. al.} {\cite{EK}}   as follows:

\be
\label{redfor}
 {\sigma_N}{\left({\phi_i}\right)}    \quad   \quad  =
 \quad  {W^\prime} {\int^x}{\sigma_{N - 1}}{\left({\phi_i}\right)}
 \quad  +  \quad   {\sum_{j = 0}^k} \, {\phi_{k - j}} \,
{\left( {{\partial}\over{\partial{t_j}}} \,
{{\Bbb R}_{(N)}^i}{\left({\underline t}\right)}\right)}
\ee
where we have defined:
\be
\label{defr}
 {{\Bbb R}_{(M)}^i}{\left({\underline t}\right)}
\quad   \stackrel{\rm def}{=}   \quad   -   \,  b(M + 1, i) \,
  {\oint}dx \, {L^{M(k + 2) + i + 1}}
\ee
The quantity  $  {{\Bbb R}_{(M)}^i}{\left({\underline t}\right)}$
   may be identified with the Gelfand-Dikki potential of the
 KdV hierarchy.  From the above definition, we can at once
 draw  the folllowing inferences.

\ben
\item    Clearly as a special case (setting all
gravitational couplings to zero) we see:
\bea
 {{\Bbb R}_{(0)}^i}{\left({\underline t}\right)}   &
\quad  =    \quad  &  -  \, {\left({1\over{i + 1}}\right)}\,
  {\oint}dx \, {L^{  i + 1}}\no\\
\label{coup}   {\impl} \,
{{\Bbb R}_{(0)}^i}{\left({\underline t}\right)}
 &  \quad  =    \quad  &  {t_{0,{k - i}}}\\
{\hbox{Similarly}}  \quad  \quad \quad   \quad
 \quad \quad \quad \quad
 {{\Bbb R}_{(1)}^i}{\left({\underline t}\right)}
 &   \quad   =    \quad   &    -   \,
 {1\over{{\left(i + 1\right)}{\left(k + i + 3\right)}}} \,
 {\oint}dx \, {L^{ k   +  i + 3}}\no\\
{\impl} \,  {{\Bbb R}_{(1)}^i}{\left({\underline t}\right)} {\gf}
     &   \quad  =    \quad   &    -   \,
  {{\partial {{\cal F}_0}}\over{\partial{t_i}}}
\eea
where ${{\cal F}_0}{\left({t_0},{t_1}, \ldots
,{t_k}\right)}$  is the free-enegy of the model
 in the absence of gravity.

\item  From the definition, we also see that
\be
 {{\Bbb R}_{(N)}^i}{\left({\underline t}\right)}  \quad =
  \quad  0  \quad \quad \quad \quad {\hbox{for}}  \, \, N \, \, <  \, \, 0.
\ee
Further we can also establish the relation
\be
{{(- 1)}^r} {{\partial^r}\over{\partial{{t_0}^r}  }} \,
  {{{\Bbb R}_{(N)}^l}{\left({\underline t}\right)}_{\gf}}
 \quad =  \quad   {{\Bbb R}_{(N - r)}^l}{\left({\underline t}\right)}
\ee
for the quantity  ${{\Bbb R}_{(N)}^i}{\left({\underline t}\right)}$
 by putting all the couplings to the gravitational descendants
to zero at the end of the day.
\een

\nd
Thus we have the very important {\it reduction formula  for
the gravitational descendants}:
 {\boldmath{
\bea
 {\forall}  \, \, \, N  \, \, {>}  \,  \, 0,   \quad \quad
 {\sigma_N}{\left({\phi_i}\right)} \,    &  =    &  \,
{W^\prime} {\int^x}{\sigma_{N - 1}}{\left({\phi_i}\right)}
 \,   +  \,    {\sum_{j = 0}^k} \, {\phi_{k - j}}
\,{\left( {{\partial}\over{\partial{t_j}}}
\, {{\Bbb R}_{(N)}^i}{\left({\underline t}\right)}\right)}\no\\
{\hbox{while for}}  \quad  N \,  =  \, 0, \quad \quad
     {\sigma_0}{\left({\phi_i}\right)} \,   &  =   &
 \, {\phi_i}  \quad   \quad \quad       {\forall} \,
 \, \,  i, \, \, 0 \, \,  {\leq}  \, \,  i  \, \,
 {\leq}  \,  \, k\\
{\hbox{and  for}}  \quad  N \,  {<}  \, 0, \quad \quad
     {\sigma_N}{\left({\phi_i}\right)} \,   &  =   &
  \, \, \,  0  \quad   \quad \quad       {\forall} \,
  \,  \,  i, \, \, \,  0 \,  \, {\leq}  \,  \, i  \,
 \,  {\leq}  \,  \, k\no
\eea
}}

\nd
Clearly from the above reduction formula, we get for
$N \, >  \, 0$,
the useful identification:
{\boldmath
\be
\label{resa}
<{\sigma_N}{\left({\phi_i}\right)}{\phi_j}>   \quad    =
  \quad     {{\partial}\over{\partial{t_j}}} \,
 {{\Bbb R}_{(N)}^i}{\left({\underline t}\right)}
\ee
}
Further, inserting the above result into the reduction
 formula for the gravitational descendants, we get the
 following alternatve form:
{\boldmath
\be
 {\forall}  \, \, \, N  \, \, {>}  \,  \, 0,
  \quad \,  {\sigma_N}{\left({\phi_i}\right)} \quad
     =      \quad
  {W^\prime} {\int^x}{\sigma_{N - 1}}{\left({\phi_i}\right)}
 \,   +  \,    {\sum_{j = 0}^k} \,
 {\phi_{k - j}} \,<{\sigma_N}{\left({\phi_i}\right)}\, {\phi_j}>
\ee
}

\bigskip

\bigskip


\section{Some important  properties  and results}
\setcounter{equation}{0}

 In this section, we shall try to obtain some useful
 conclusions which follow readily from   our constructions.
 Some of these results (like the {\it contact algebra})
 are already known in the context of discussions of $2d$
 topological gravity based on the path-integral or other
 approaches. We shall rederive these results and also
extend and generalize some of them.  We will be able to
 obtain an interesting hierarchy of differential equations
(which are supposed to be the generalization of the
multi-contact term algebra).  We shall also   identify an
 {\it operator correspondence}  which,   we expect,  does
have deeper mathematical implications as similar operations
 also appear in an entirely different context (in the
mathematical works of {\it Saito} on {\it higher residue
 pairing}).    Finally we shall obtain an important recursive
 differential equation for the
$ {{\Bbb R}_{(N)}^l}{\left({\underline t}\right)}$,
 which incorporates the multiplicative chiral ring
property of the primary fields  of the theory. This
 relation is reminiscent of the {\it flatness} criterion
 on the space of couplings.

\subsection{Contact Algebra}

Let us now try to obtain some equations (in the large phase
space) which follow directly from the reduction formula for
the gravitational descendants. In  the analysis of $2d$
gravity (from other standpoints), it is well known that
 there are {\it contact terms} in the expressions for the
correlation functions. We shall now obtain the equivalent
 expressions for these contact terms based on our approach.

\ben

\item  From the  expression:
$$
{\left({{\sigma_N}{\left({\phi_i}\right)}\over{W^\prime}}\right)}
    \quad  =    \quad   {\int^x}{\sigma_{N - 1}}{\left({\phi_i}\right)}
 \quad  +  \quad   {\sum_{j = 0}^k} \,
 {\left({{\phi_{k - j}}\over{W^\prime}}\right)} \
,{\left( {{\partial}\over{\partial{t_j}}} \,
 {{\Bbb R}_{(N)}^i}{\left({\underline t}\right)}\right)}$$
we have the important result:
{\boldmath
\be
\label{propa1}
{\partial_x} {{\left[\,
{\left({{\sigma_N}{\left({\phi_i}\right)}\over{W^\prime}}\right)}\,
\right]}_+}  \quad  =  \quad  {\sigma_{N - 1}}{\left(\phi_i\right)}
\ee
}
Thus setting $N \, =  \, 0$ in the above equation we get:
\be
\label{propa2}
 {\partial_x} {{\left[\,
{\left({{\sigma_0}{\left({\phi_i}\right)}\over{W^\prime}}\right)}\, \right]}_+}
 \quad  {\equiv}   \quad    {\partial_x}
{{\left[\,{\left({ {\phi_i} \over{W^\prime}}\right)}\,
 \right]}_+}  \quad  =  \, 0
\ee
Using the definitions it is quite simple to derive
the following set of first-order differential equations
governing the $t_{N,i}$ dependence of the descendant
fields  ${\sigma_N}{\left({\phi_i}\right)}$
{\boldmath
\be
\label{contact}
{{\partial \, {\sigma_N}{\left({\phi_i}\right)}
}\over{\partial{t_{M,j}}}}  \,    =     \,
 {{\partial \,{\sigma_M}{\left({\phi_j}\right)}
}\over{\partial{t_{N,i}}}}   \,  =  \,     - \,
 {\partial_x} {{\left[\,{\left({  {\sigma_N}
{\left({\phi_i}\right)}\,
{\sigma_M}{\left({\phi_j}\right)} \over{W^\prime}}\right)}\,
\right]}_+}
\ee
}
These equations are {\it integrable}, and in
fact their integrability may be verified by
 noting that an {\it explicit} solution to the
above system is given by:

\be
{\sigma_N}{\left({\phi_i}\right)}  \quad {\sim} \quad
  - {1\over{N(k + 2) + i}}
{\Bigl[{\left( - W\right)}^{N(k + 2) + i}\Bigr]}^{\prime}_{+}
\ee
where the primes refer to derivatives with respect to $x$.

An operator ${\sigma_N}{\left({\phi_i}\right)}$ inserted
in a  correlation function creates a puncture and thus
introduces a moduli that must then be integrated over.
Thus in effect ${\sigma_N}{\left({\phi_i}\right)}$ is
represented by an integration of its 2-form partner.
This integration receives a special contribution whenever
 ${\sigma_N}{\left({\phi_i}\right)}$ approaches some other
operator ${\sigma_M}{\left({\phi_j}\right)}$  on the surface.
This special contribution is what we call the contact term.
  Thus in the language of $2d$ gravity the term
$C{\left(N,i;M,j\right)}$ is identified as a {\it contact}
term created by the {\it collision} of two  gravitational
descendant operators, where

$$C{\left(N,i;M,j\right)}  {\dfn}
{\Bigl[{{{\sigma_N}{\left({\phi_i}\right)}
 {\sigma_M}{\left({\phi_j}\right)}}\over{W^\prime}}\Bigr]}^{\prime}_{+}
$$
These contact terms are of great significance
in obtaining the expressions for the $N$-point ($N \, >
 \, 3$) correlation functions of the theory as it is
the presence of these contact terms that  render
these generic  $N$-point ($N \, > \, 3$) correlation
functions {\it symmetric} under the interchange of of
its arguments.   The contact term $C{\left(N,i;M,j\right)}$
expresses the gravitational {\it dressing} of the
descendant field ${\sigma_N}{\left({\phi_i}\right)}$
by the couplings $t_{M,j}$ and we note the nice {\it duality}
 that this is the same as the
 gravitational {\it dressing}  of the descendant field
${\sigma_M}{\left({\phi_j}\right)}$ by the couplings $t_{N,i}$.

\bigskip

{}From (\ref{contact}),  we may conclude the  following
by  setting  different specific   values of $N, M$:

\bigskip

\framebox[1in]{$M = N = 0$}  \, :  \quad  In this case, we get
\bea
\label{propbb}
{{\partial \, {\sigma_0}{\left({\phi_i}\right)}
}\over{\partial{t_{0,j}}}}  \quad  &  =  &    \quad
 - \,  {\partial_x} {{\left[\,{\left({
{\sigma_0}{\left({\phi_i}\right)}\,
 {\sigma_0}{\left({\phi_j}\right)}
 \over{W^\prime}}\right)}\, \right]}_+}\no\\
{i.e.,}  \quad  \quad  {{\partial \,
 { {\phi_i}}  }\over{\partial{t_{j}}}}
 \quad  &  =    &  \quad   - \,
 {\partial_x} {{\left[\,{\left({  {{\phi_i}}\,
  {{\phi_j}}   \over{W^\prime}}\right)}\, \right]}_+}
\eea
exactly the same as is known  in the gravity-free case.
  Further, the above equation is of great significance
 as it imposes the condition of {\it flatness} on the
perturbing coordinates---  the flat-coordinates  are
thus the solutions to the above equation.

\bigskip

\framebox[.6in]{$M = 0$}  \, :  \quad  In this case,
we get
$$
 {{\partial \, {\sigma_N}{\left({\phi_i}\right)}
 }\over{\partial{t_{j}}}}  \,    =    \,    {{\partial \,
   {{\phi_j}} }\over{\partial{t_{N,i}}}}  \,    =     - \,
  {\partial_x} {{\left[\,{\left({
{\sigma_N}{\left({\phi_i}\right)}\,  {{\phi_j}}
 \over{W^\prime}}\right)}\, \right]}_+}
$$
which tells us that:
\bea
\label{propc}
{{\partial \,    {{\phi_j}} }\over{\partial{t_{N,i}}}}
 \quad  &  =    &  \quad   - \,  {\partial_x}
 {{\left[\,{\left({  {\sigma_N}{\left({\phi_i}\right)}\,
 {{\phi_j}}   \over{W^\prime}}\right)}\, \right]}_+}\\
&  {\ne}  &   \quad 0,   \quad \quad \quad {\hbox{in general}}.\no
\eea
Thus {\it  the chiral primaries acquire non-trivial dependences
(at higher order) on the gravitational couplings -- that
is they get `gravitationally dressed'}.  As a consequence,
the correlation functions involving the chiral primaries
will  {\it depend} on the couplings in the large phase space {\cite{DVV1}}.

Setting  $j \, =  \, 0$ in the above eq.(\ref{propc}),
and using the result (\ref{propa1})  gives we get  the
important relation:
{\boldmath
\be
 {{\partial \, {\sigma_N}{\left({\phi_i}\right)}
}\over{\partial{t_{0}}}}  \quad  =   \quad  -  \,  \,
 {\sigma_{N  -  1}}{\left({\phi_i}\right)}
\ee
}
We may note here that the operator ${t_0}$ couples to the
{\it puncture operator}  in our expression for the perturbed
action.  In terms of path-integrals, the operator $ {\left( - \,
 {{\partial}\over{\partial{t_0}}}\right)}$  correspondes to
${\bf P}$ operator insertions, and hence the above equation
gives us an useful reduction formula for expressing  correlation
functions of $ N^{th}$- gravitational descendant  in terms of
lower ({\it i.e.}, $(N - 1)^{th}$ and so on) descendant fields.

\item   Using the above relations recursively we can also
obtain the following important hierarchy of equations:
\bea
{{\partial \, {\sigma_N}{\left({\phi_i}\right)}
 }\over{\partial{t_{M,j}}}}  \quad &    =  &   - \,
  {\partial_x}\,  {{\left[\,{\left({
 {\sigma_N}{\left({\phi_i}\right)}\,
 {\sigma_M}{\left({\phi_j}\right)}
  \over{W^\prime}}\right)}\, \right]}_+}\no\\
{{{\partial^2} \, {\sigma_N}{\left({\phi_i}\right)}
 }\over{{\partial{t_{M,j}}}\, {\partial{t_{L,s}}}
   }}  \quad &    =  &    \, \,   {\partial^{2}_x} \,
 {{\left[\,{\left({  {\sigma_N}{\left({\phi_i}\right)}\,
  {\sigma_L}{\left({\phi_s}\right)}\,
 {\sigma_M}{\left({\phi_j}\right)}
 \over{{W^\prime}^2}}\right)}\, \right]}_+}\no\\
{{{\partial^3} \, {\sigma_N}{\left({\phi_i}\right)}
 }\over{{\partial{t_{M,j}}}\, {\partial{t_{L,s}}} \,
 {\partial{t_{P,l}}}   }}  \quad &    =  &    -  \,
 {\partial^{3}_x} \, {{\left[\,{\left({
  {\sigma_N}{\left({\phi_i}\right)}\,
  {\sigma_L}{\left({\phi_s}\right)}\,
  {\sigma_P}{\left({\phi_l}\right)}\,
    {\sigma_M}{\left({\phi_j}\right)}
  \over{{W^\prime}^3}}\right)}\, \right]}_+}\\
&  {\vdots}   &    \quad  \no\\
&  {\vdots}   &    \quad  \no\\
&  {\vdots}   &    \quad  \no
\eea
Or {\it  equivalently} the hierarchy:
\bea
{{{\partial^2} \, W  }\over{\partial{t_{M,j}}
 \,  {\partial{t_{N,i}}} }}  \quad &    =  &
 \,  {\partial_x}\,  {{\left[\,{\left({
 {\sigma_N}{\left({\phi_i}\right)}\,
{\sigma_M}{\left({\phi_j}\right)}
 \over{W^\prime}}\right)}\, \right]}_+}\no\\
{{{\partial^3} \, W  }\over{{\partial{t_{M,j}}}\,
   {\partial{t_{L,s}}}\,    {\partial{t_{N,i}}}   }}
  \quad &    =  &   -  \,  {\partial^{2}_x} \,
 {{\left[\,{\left({  {\sigma_N}{\left({\phi_i}\right)}\,
 {\sigma_L}{\left({\phi_s}\right)}\,
 {\sigma_M}{\left({\phi_j}\right)}
  \over{{W^\prime}^2}}\right)}\, \right]}_+}\no\\
{{{\partial^4} \, W  }\over{{\partial{t_{M,j}}}\,
    {\partial{t_{L,s}}} \, {\partial{t_{P,l}}}\,
 {\partial{t_{N,i}}}    }}  \quad &    =  &
  \,  {\partial^{3}_x} \, {{\left[\,{\left({
  {\sigma_N}{\left({\phi_i}\right)}\,
 {\sigma_L}{\left({\phi_s}\right)}\,
 {\sigma_P}{\left({\phi_l}\right)}\,
     {\sigma_M}{\left({\phi_j}\right)}
 \over{{W^\prime}^3}}\right)}\, \right]}_+}\\
&  {\vdots}   &    \quad  \no\\
&  {\vdots}   &    \quad  \no\\
&  {\vdots}   &    \quad  \no
\eea

\nd
Looking at the above sequence of non-linear
 differential equations, we can can make
 the following observation about the
{\it operator equivalence}, which may be symbolically expressed as:
\be
\label{opeq}
{\prod_{p = 1}^{n + 1}} \,
{\left({{\partial}\over{\partial {t_{M,{i_p}}}}}\right)}
 \quad {\longleftrightarrow} \quad   {\left( - 1\right)}^{n + 1}
\, {{\partial_x}^n} \, {{\left[{1\over{{W^\prime}^n}} \,
 {\left( \cdots\right)}\right]}_+}  \quad \quad \quad
 (n = 0, 1, \ldots ,k)
\ee
We think  that the above  observation might have deeper
   mathematical  implications in terms of Saito's {\it higher
 residue pairing} {\cite{{EK},{LOS},{SAI}}}, but this needs
 to be investigated further.

\een

\subsection {Recursive Property for  ${{R}_{(N)}^l}
{\left({\underline t}\right)}$     }

We shall now obtain an interesting property for the
${{\Bbb R}_{(N)}^l}{\left({\underline t}\right)}$
by exploiting the chiral ring structure of the matter
sector of the theory. We shall obtain a recursive
second-order differential  equation with respect
to the parameters in the small phase space for
${{\Bbb R}_{(N)}^l}{\left({\underline t}\right)}$.
 It is interesting to note that in this equation all
the differential operators act only on the small phase
space. We shall see in a subsequent section that all
the coordinates  in the large phase space can be
eventually expressed in terms of  those on the small
phase space, and thus such a relation is not totally
unexpected. We shall evoke the  multiplicative properties
of the chiral ring of the primary fiels in the spectrum
of the theory (the ring structure is still preserved even
though it is no longer a nilpotent ring) to obtain this
desired relation.

Beginning from the definition ({\ref{defr}}) we have:
\bea
{{\partial^2}\over{{\partial{t_i}}{\partial{t_j}}}}
{{\Bbb R}_{(N)}^l}{\left({\underline t}\right)}
  &  \quad    =    \quad   &   - \,  b(N - 1,l) \,
  {\oint}dx \, {L^{(N - 2)(k + 2) + l + 1}} \,
 {\left({\phi_i}{\phi_j}\right)}\no\\
&  \quad  &   \quad  \quad  \quad   \quad
\quad
 \quad  \quad  +  \,  \,  b(N,l) \, {\oint}dx \,
 {L^{(N - 1)(k + 2) + l + 1}} \,
{\left({{\partial{\phi_i}}\over{\partial{t_j}}}\right)}\no\\
&  \quad    =    \quad   &   - \,
{\sum_m} \,\,  {{{\cal C}_{ij}} ^m} \,
 {\left({{\partial}\over{\partial{t_m}}}
{{\Bbb R}_{(N - 1)}^l}{\left({\underline t}\right)}\right)}\no\\
&  \quad  &   \quad  \quad  \quad   \quad
 \quad   \quad  \quad  +  \, \,  b(N,l) \,
 {\oint}dx \, {L^{(N - 1)(k + 2) + l + 1}} \,
 {\left({{\partial{\phi_i}}\over{\partial{t_j}}}\right)}\no
\eea
where we have used the multiplicative  chiral
ring algebra   in the form:
$${\phi_i} \, {\phi_j}  \quad = \quad {\sum_m}\,
 {{{\cal C}_{ij}}^m}\, {\phi_m}$$
with  ${{{\cal C}_{ij}}^m} \,  \, {\equiv}  \, \,
  {{{\cal C}_{ij}}^m}{\left({t_0}, {t_1}, {t_2},
\ldots  ,{t_k}\right)}$  being the  structure
constants of the chiral ring algebra of our $N = 2, \,
{A_{k + 1}}$ model.

\nd
Thus we may write the above equation in the form:
\be
b(N,l) \, {\oint}dx \, {L^{(N - 1)(k + 2) + l + 1}} \,
 {\left({{\partial{\phi_i}}\over{\partial{t_j}}}\right)}
   =   {{\partial^2}\over{{\partial{t_i}}{\partial{t_j}}}}
 {{\Bbb R}_{(N)}^l}{\left({\underline t}\right)}  \,
 +  \,  {\sum_m} \,\,  {{{\cal C}_{ij}} ^m} \,
 {\left({{\partial}\over{\partial{t_m}}}
{{\Bbb R}_{(N - 1)}^l}{\left({\underline t}\right)}\right)}
\ee
An interesting fact of life  is that  the left-hand
side  expression of the above equation may be shown
 actually to {\it vanish} exactly  ({\ref{appc}})
(for proof see the appendix), {\it i.e.},
\be
\label{vanish}
 {\oint}dx \, {L^{(N - 1)(k + 2) + l + 1}} \,
{\left({{\partial{\phi_i}}\over{\partial{t_j}}}\right)}
 \quad  = \quad  0
\ee
Leaving us with  the following important property:

\medskip
{\boldmath {
\be
\label{propr}
{{\partial^2}\over{{\partial{t_i}}{\partial{t_j}}}}
{{\Bbb R}_{(N)}^l}{\left({\underline t}\right)}  \, \,
   +  \, \,  {\sum_m} \,\,  {{{\cal C}_{ij}} ^m} \,
{\left({{\partial}\over{\partial{t_m}}}
{{\Bbb R}_{(N - 1)}^l}{\left({\underline t}\right)}\right)}
 \quad  =  \quad 0
\ee
}}

\nd
The above gives us an useful recursive, second-order
differential equation for the
${{\Bbb R}_{(N)}^l}{\left({\underline t}\right)}$,
 which also introduces the structure constants of
the chiral ring into the picture. Let us then analyze
 some special cases of the above equation.

Before that in passing, we can make the following
interesting {\it observation} here. Let us  {\it define}
 a quantity ${\Psi}^{(l)} {\left({\underline t} \right)}$ by

\be
\label{defs}
{\Psi^{(l)}} {\left({\underline t} \right)}    \,
  {\dfn}  \,  {\sum_{N = - \infty}^{N = \infty}} \,
{{\Bbb R}_{(N)}^l}{\left({\underline t}\right)}
 \quad  =  \quad   {\sum_{N = 0}^{N = \infty}} \,
{{\Bbb R}_{(N)}^l}{\left({\underline t}\right)}
\ee
the equality of the two summations being obvious because
$ {{\Bbb R}_{(N)}^l}{\left({\underline t}\right)} \, = \,
 0,$ for $N \, < \, 0$. Now if we sum both sides of the
equation ({\ref {propr}}) over $N = -{\infty}$ to ${\infty}$,
  we have the following relation for  ${\Psi}^{(l)}
{\left({\underline t} \right)}$:

\be
{\biggl[  \, {{\partial^2}\over{{\partial{t_i}}{\partial{t_j}}}}
 \, \,   +  \, \,  {\sum_m} \,\,  {{{\cal C}_{ij}} ^m} \,
 {\left({{\partial}\over{\partial{t_m}}}    \right)}  \,
 \biggr]} \, {\Psi}^{(l)} {\left({\underline t} \right)}
     \quad  =  \quad 0
\ee
The differential operator within  the square braces
above can be easily identified with the familiar covariant
 derivatives and the above equation can be interpreted
as a {\it flatness} criterion  in the small phase space
for the quantity
${\Psi}^{(l)} {\left({\underline t} \right)}$.

\medskip

\nd
{\bf Special case: } \quad  If we set all {\it gravitational
couplings to zero}, as a special case, we see that the above
 equation may be simplified further as follows.
Using the equation ({\ref{proprr}})  (proved later)\, \,
 ${{\partial}\over{\partial{t_0}}}
{{\Bbb R}_{(N)}^l}{\left({\underline t}\right)}
\quad  =  \quad   - \,
{{\Bbb R}_{(N - 1)}^l}{\left({\underline t}\right)}$
 in the gravity-free case, we can also rewrite the
above equation in the alternative form:
{\boldmath
\bea
{{\grd}_{ij}} \,
{{{\Bbb R}_{(N)}^l}{\left({\underline t}\right)}_{\gf}}
 \quad  &  =    &  \quad 0\\
{\hbox{where}}  \quad  \quad  {{\grd}_{ij}}  \quad  &
  {\equiv}  &    \quad  {{\grd}_{ij}}{\left({t_0}, {t_1},
 {t_2}, {\ldots}  , {t_k}\right)}\no\\
&  {\dfn}  & {\biggl[ {{\partial^2}\over{\partial{t_i}}{\partial{t_j}}}
  \, -  \, {\sum_{m = 0}^k}\, {{{\cal C}_{ij}}^m}
\, {{\partial^2}\over{\partial{t_0}}{\partial{t_m}}}\biggr]}\no
\eea
}

\bd
\item [ Corollary:]  \, \,   Setting $N \, =  \, 1$  and also
all gravitational couplings to zero at the end of the day, in
the previous equation ({\ref{propr}}) we get :
\bea
{{\partial^2}\over{{\partial{t_i}}{\partial{t_j}}}}
{{\Bbb R}_{(1)}^l}{\left({\underline t}\right)}  \, \,
+  \, \,  {\sum_m} \,\,  {{{\cal C}_{ij}} ^m} \,
 {\left({{\partial}\over{\partial{t_m}}}
{{\Bbb R}_{(0)}^l}{\left({\underline t}\right)}\right)}
 &    =  &  0\no\\
{\impl} \quad  {{\partial^2}\over{{\partial{t_i}}{\partial{t_j}}}}
{{\Bbb R}_{(1)}^l}{\left({\underline t}\right)}{\gf}
 \, \,   +  \, \,  {\sum_m} \,\,  {{{\cal C}_{ij}} ^m}
 \, {\left({{\partial}\over{\partial{t_m}}}{{\Bbb R}_{(0)}^l}
{\left({\underline t}\right)}\right)}{\gf}  &    =  &  0\no\\
{\hbox{Hence}},   \quad  \quad
{{\partial^2}\over{{\partial{t_i}}{\partial{t_j}}}}
 {{\Bbb R}_{(1)}^l}{\left({\underline t}\right)}{\gf}
&   =   &   - \,  {{{\cal C}_{ij}} ^{k - l}}\no\\
&  =  &   - \, {{c_{ij}}^l}\no
\eea
where ${{c_{ij}}^l} \,  =  \, {{c_{ij}}^l}{\left(
{t_0}, {t_2}, \ldots  ,{t_k}\right)}$  is related to
 the 3-point correlation function of the {\it gravity-free}
model via the rule: ${{c_{ij}}^l}  \, \, =  \,  \, {c_{ijm}}
 \, {\eta^{lm}_{(0)}}$. The above result also confirms our
 previous identification:
$$
{{\Bbb R}_{(1)}^i}{\left({\underline t}\right)}{\gf}
 \quad  =    \quad       -   \,  {{\partial
 {{\cal F}_0}}\over{\partial{t_i}}}
$$
\ed

\subsection{Perturbed Correlation Functions }

Finally, using the expressions for the gravitational
descendants we see that the 2- and 3-point  correlation
 functions  of the theory  can  be easily calculated as
follows. Further knowing these functions, all the
higher order correlation functions of the theory
can also be easily obtained.
\ben
\item The 2-point function is given by:
\bea
<{\sigma_M}{\left({\phi_i}\right)}{\sigma_N}{\left({\phi_j}\right)}>
 \, \,  &      =  &     {\Biggl[ \, {\sum_{p = 0}^k} \,
  {\sum_{q = 0}^k} \,\, {\left( {{\partial}\over{\partial{t_p}}} \,
 {{\Bbb R}_{(M)}^i}{\left({\underline t}\right)}\right)}
   \,{\left( {{\partial}\over{\partial{t_q}}} \,
 {{\Bbb R}_{(N)}^j}{\left({\underline t}\right)}\right)}\Biggr]}
<{\phi_{k - p}}{\phi_{k - q}}>\no\\
&  =  &   {\Biggl[ \, {\sum_{p = 0}^k} \,
 {\sum_{q = 0}^k} \,\, {\left( {{\partial}\over{\partial{t_p}}} \,
 {{\Bbb R}_{(M)}^i}{\left({\underline t}\right)}\right)}
     \,{\left( {{\partial}\over{\partial{t_q}}} \,
 {{\Bbb R}_{(N)}^j}{\left({\underline t}\right)}\right)}\Biggr]}
 {\delta_{p +  q , k}} \no
\eea

\nd
Thus we have:
{\boldmath
\be
<{\sigma_M}{\left({\phi_i}\right)}
{\sigma_N}{\left({\phi_j}\right)}>
\quad   =  \quad     {\Biggl[ \, {\sum_{p = 0}^k} \,
   \,\, {\left( {{\partial}\over{\partial{t_p}}} \,
 {{\Bbb R}_{(M)}^i}{\left({\underline t}\right)}\right)}
       \,{\left( {{\partial}\over{\partial{t_{k - p}}}}
 \, {{\Bbb R}_{(N)}^j}{\left({\underline t}\right)}\right)}\Biggr]}
\ee
}

\medskip
\nd
{\bf  {Factorization}}:  \,  It is known from general
 theorems of topological theories that the correlation
 functions satisfy  the {\it factorization} hypothesis. We can easily check
this explicitly from our construction as follows.
  The 2-point function satisfies the important factorization
 property as can now be easily verified. Using the result
({\ref{resa}}), we can rewrite the expression for the 2-point function as:
\bea
<{\sigma_M}{\left({\phi_i}\right)}{\sigma_N}{\left({\phi_j}\right)}>
\quad  &   =  &     {\Biggl[ \, {\sum_{p = 0}^k} \,
  \,\, {\left( {{\partial}\over{\partial{t_p}}} \,
 {{\Bbb R}_{(M)}^i}{\left({\underline t}\right)}\right)}
     \,{\left( {{\partial}\over{\partial{t_{k - p}}}} \,
 {{\Bbb R}_{(N)}^j}{\left({\underline t}\right)}\right)}\Biggr]}\no\\
&  =  &  {\sum_{p = 0}^k} \, { <{\sigma_M}{\left({\phi_i}\right)}\,
 {\phi_p}> \, <{\sigma_N}{\left({\phi_j}\right)} \,
 {\phi_{k -p}}> }\no
\eea
Since  ${\phi^p}  \, =   \, {\eta^{pm}_{(0)}}{\phi_m}
  \, =  \,  {\delta^{p + m, k}}{\phi_m}  \, =  \,
 {\phi_{k - p}}$,  we have the nice  {\it factorization} property:
\be
<{\sigma_M}{\left({\phi_i}\right)}{\sigma_N}{\left({\phi_j}\right)}>
 \quad   =   \quad   {\sum_{m,l = 0}^k} \,
{ <{\sigma_M}{\left({\phi_i}\right)}\, {\phi_m}> \,
{\eta^{ml}_{(0)}}\, <   {\phi_{l}}  \,
 {\sigma_N}{\left({\phi_j}\right)} > }
\ee

\item   Similarly, the expression for the
3-point function becomes:
\bea
<{\sigma_M}{\left({\phi_i}\right)}{\sigma_N}
{\left({\phi_j}\right)}{\sigma_P}{\left({\phi_l}\right)}>
 \,   &    =      &   {\sum_{ {\lbrace  {p,q,r}\rbrace}  = 0}^k}
 {\Biggl[   \,{\left( {{\partial   \,
 {{\Bbb R}_{(M)}^i}{\left({\underline t}\right)}
 }\over{\partial{t_{k - p}}}} \,  \right)} \,
 \,{\left( {{\partial    \,
 {{\Bbb R}_{(N)}^j}{\left({\underline t}\right)}
 }\over{\partial{t_{k - q}}}} \,   \right)}   \,
 \,{\left( {{\partial     \,
 {{\Bbb R}_{(P)}^l}{\left({\underline t}\right)}
   }\over{\partial{t_{k - r}}}} \,
   \right)}\Biggr]}\no\\
&   \quad   &   \quad  \quad  \quad
 \quad  \quad  \quad  \quad   \quad   \quad
  \quad   \quad  \quad  \quad  \quad  \quad
\quad     {\cdot} \, <{\phi_{p}} {\phi_{q}}
 {\phi_{r}}>\no\\
 &    =    &   {\sum_{{\lbrace{p,q,r}\rbrace}
 = 0}^k}  {\Biggl[   \,{\left( {{\partial   \,
  {{\Bbb R}_{(M)}^i}{\left({\underline t}\right)}
    }\over{\partial{t_{k - p}}}} \,  \right)} \,
 \,{\left( {{\partial    \,
  {{\Bbb R}_{(N)}^j}{\left({\underline t}\right)}
 }\over{\partial{t_{k - q}}}} \,   \right)}
 \, \,{\left( {{\partial     \,
  {{\Bbb R}_{(P)}^l}{\left({\underline t}\right)}
  }\over{\partial{t_{k - r}}}} \,
  \right)}\Biggr]}\no\\
&   \quad   &    \quad  \quad  \quad
 \quad  \quad  \quad  \quad  \quad \quad
 {\cdot} \,
 {{{\partial_p}{\partial_q}}\over{{\left(r
+ 1\right)}{\left(k + r + 3\right)}}} \,
{\oint}dx  \, {L^{k + r + 3}}\no
\eea

\een

\nd
Therefore the expression for the 3-point
function is:
{\boldmath
$$<{\sigma_M}{\left({\phi_i}\right)}
{\sigma_N}{\left({\phi_j}\right)}{\sigma_P}
{\left({\phi_l}\right)}> \quad =  \quad
 \quad \quad \quad   \quad \quad \quad
 \quad  \quad \quad \quad \quad   \quad
 \quad \quad  \quad   \quad \quad \quad
 \quad  \quad \quad \quad \quad \quad
$$
\be
\label{eqc}    \quad \quad \quad  -  \,
 {\sum_{{\lbrace{p,q,r}\rbrace}  = 0}^k}
 {\Biggl[   \,{\left( {{\partial   \,
  {{\Bbb R}_{(M)}^i}{\left({\underline t}\right)}
   }\over{\partial{t_{k - p}}}} \,
 \right)} {\left( {{\partial    \,
 {{\Bbb R}_{(N)}^j}{\left({\underline t}\right)}
 }\over{\partial{t_{k - q}}}} \,
   \right)}   {\left( {{\partial     \,
  {{\Bbb R}_{(P)}^l}{\left({\underline t}\right)}
  }\over{\partial{t_{k - r}}}} \,
   \right)}\Biggr]}  {\cdot}
  {\left({{\partial^2}\over{{\partial{t_p}}{\partial{t_q}}}}
\, {{\Bbb R}_{(1)}^r}{\left({\underline t}\right)}\right)}
\ee
}
Setting the indices $M = N = P = 0$
in the above then gives us back the expressions
well known {\cite{DVV1}} for the gravity-free theory.
\bigskip


\section { Recursion Relations  for Correlation Functions  }
\setcounter{equation}{0}

In this section, we shall prove some  exact relations
 for correlation functions in our gravity-coupled model.
 The nice thing about these relations is that these
 will be valid in the {\it entire} phase space of our
theory  ({\it i.e.}, will be valid on the large phase
space as well). When reformulated in the language of
 path-integrals, these relations   then  become the
generalized Ward identitites for the theory. Examples
of such relations (like the puncture equation) valid
on the  {\it small phase space} have  been used in the
  discussions of topological gravity {\cite{DVV2}},
{\cite{VV}}, and we now try to extend such results to
the {\it large phase space}. As a by-product, we also
rederive,  in an alternative fashion,  the known results
 (on the {\it small phase space}). We show the calculations
 in some  details as some highly non-trivial manipulations
 are involved in the intermediate stages.

\subsection {Ward Identity}

{}From the  definition {\cite{VA}} of correlation functions
 in {\it genus}-  0, we get:
\bea
{{\partial}\over{\partial{t_{N,i}}}}
 <{\sigma_M}{\left({\phi_j}\right)}>  \,   &  =
&  {{\partial}\over{\partial{t_{N,i}}}}\,   {\oint}dx \,
{\left({{{\sigma_M}{\left({\phi_j}\right)}}\over{W^\prime}}\right)}\no\\
 & =     &      - \, \,    {\oint}dx \,
{\left({1\over{W^\prime}} \right)}\, {\partial_x}
{{\left[{{{\sigma_M}{\left({\phi_j}\right)}\,
{\sigma_N}{\left({\phi_i}\right)}}\over{W^\prime}}\right]}}\no\\
& \quad     &   \quad  \quad      -  \,
  {\oint}dx \,
 { {{\sigma_N}{\left({\phi_i}\right)}}\over{W^\prime}}
  \,     {\partial_x}
 {\left[ {{{{\sigma_M}{\left({\phi_j}\right)}}\over{W^{{\prime}}}}}   \right]}
 \,    -  \,   {\oint}dx \,
  { {  {{\sigma_N}{\left({\phi_i}\right)}} {\sigma_M}{\left({\phi_j}\right)}
}\over{W^\prime}}    \,
 {\partial_x} {\left( {{{  1  }\over{W^{{\prime}}}}}   \right)}\no\\
&  \quad &  \quad
 ({\hbox{where we have used  eq.
(\ref{propa1})  to replace the first term, }}\no\\
&  \quad  &  \quad   \quad  \quad
  {\hbox{ and have also   subsequently
 neglected the + subscript}})    \no\\
& =   &     -  \, \,   {\oint}dx \,
 {\left( { {{\sigma_N}{\left({\phi_i}\right)}}\over{W^\prime}}\right)}
   \,     {\partial_x}
{\left[ {{{{\sigma_M}{\left({\phi_j}\right)}}\over{W^{{\prime}}}}}
  \right]}     \\
&  =  &   -   <{{\sigma_N}{\left({\phi_i}\right)}}\,
  {{\sigma_{M - 1} }{\left({\phi_j}\right)}}> \,   +
  \, \,    {\sum_{l = 0}^k}  {\left( {{\partial}\over{\partial{t_l}}}
  {{\Bbb R}_{(M)}^j}{\left({\underline t}\right)}\right)}
   \, {\oint}  {\left( { {\phi_{k - l}}\over{W^\prime}}\right)}  \,
  {{\sigma_{N - 1}}{\left({\phi_i}\right)}}\no\\
&  =  &   - \,  <{{\sigma_N}{\left({\phi_i}\right)}}\,
  {{\sigma_{M - 1} }{\left({\phi_j}\right)}}>
   \,  +  \,  <{{\sigma_{N - 1}}{\left({\phi_i}\right)}}\,
   {{\sigma_{M } }{\left({\phi_j}\right)}}>\no
\eea

\nd
In the intermediate steps, we have often traded
 the ${[ \, {\ldots} \, ]}_+$  contribution of
certain terms in favour of the entire ({\it i.e.},
 without the +subscript) term. However,  in  reality,
 no harm is done   as the contribution from the
terms thus added or discarded   actually {\it vanish}.
 Thus for example in one of the  intermediate stages of
 our calculation, we have thrown off the term:  \, \,
       $  {\oint}dx \, {\left({1\over{W^\prime}} \right)}\,
 {\partial_x} {{\left[{{{\sigma_M}{\left({\phi_j}\right)}\,
 {\sigma_N}{\left({\phi_i}\right)}}\over{W^\prime}}\right]}_-}$.
\, However, since \, ${1\over{W^\prime}} \  {\sim} \, {1\over{x}}
 \, +  \, {1\over{x^2}}  \, +  \, {\ld}$, \, one can easily
 see that the contribution from the above term {\it vanishes}.
Similarly, we have also neglected the term: \,  $  {\oint}dx \,
   {\left( { {\phi_{k - l}}\over{W^\prime}}\right)}  \,
 {\partial_x}{{\left[ {\left( { {{\sigma_N}{\left({\phi_i}\right)}}
\over{W^\prime}}\right)}\right]}_-}$.  Once again,  since
$ {\left( { {\phi_{k - l}}\over{W^\prime}}\right)}  \,
 \sim  \, {1\over{x}}
 \, + {\ld}$  \, and this implies that  the contribution
from this term also {\it  vanishes}.  Hence we have the
  following important equation:

{\boldmath
\be
\label{opun}
{{\partial}\over{\partial{t_{N,i}}}}
<{\sigma_M}{\left({\phi_j}\right)}>
  \,    \,   =   \,    - \,  <{{\sigma_{M - 1}}{\left({\phi_j}\right)}}\,
  {{\sigma_{N} }{\left({\phi_i}\right)}}>
   \,  +  \,
  <{{\sigma_{M }}{\left({\phi_j}\right)}}\,
  {{\sigma_{N - 1} }{\left({\phi_i}\right)}}>
\ee
}
\nd
We note that in  the path-integral formulation,
 the operation

of ${\left( - {{\partial}\over{\partial{t_{N,i}}}}\right)}$\,
corresponds, in the {\it lowest} order,
 to the  \,
${\sigma_N}{\left({\phi_i}\right)}$ \,
 operator insertions.   The above
 equation shows how the expectaion value
of  ${\sigma_M}{\left({\phi_j}\right)}$
depends on the coupling $t_{N,i}$.
{}From the above equation, we may now make
the following conclusions.

\subsubsection{Some  Special Cases}

Setting specific values to the indices $N, M, i,
j$ in the above   equation  ({\ref{opun}})
(and always remembering the constraint: \,
 ${\sigma_{-1}}{\left({\phi_j}\right)} \quad
{\equiv}   \quad   0$),  we get the following conclusions:

\ben

\item   \framebox[.6in]{$N = 0$} :  In  this case
the  above equation tells us:
\bea
{{\partial}\over{\partial{t_{0,i}}}}\,
 <{\sigma_M}{\left({\phi_j}\right)}>
  \quad   &  =   &  \quad   - \,
<{{\sigma_{M - 1}}{\left({\phi_j}\right)}}\,
  {{\sigma_{0} }{\left({\phi_i}\right)}}>\no\\
{\impl}  \quad {{\partial}\over{\partial{t_{i}}}}\,
 <{\sigma_M}{\left({\phi_j}\right)}>
  \quad   &  =   &  \quad   - \,
 <{{\sigma_{M - 1}}{\left({\phi_j}\right)}}\,
    {\phi_i}>\no\\
&  =  &    \quad   - \,{ {{\partial}\over{\partial{t_i}}}}
 \, {{\Bbb R}_{(M - 1)}^j}{\left({\underline t}\right)}\no
\eea
Thus we once again arrive at the equation:
$$
{ {{\partial}\over{\partial{t_i}}}}
 \, {{\Bbb R}_{(N)}^j}{\left({\underline t}\right)}
 \quad =  \quad  <{{\sigma_{N}}{\left({\phi_j}\right)}}\,
    {\phi_i}>
$$
which is exactly the same ({\ref{resa}}) as derived earlier.

 We also have the useful equation:
{\boldmath
\be
\label{ida}
 {{\partial}\over{\partial{t_{i}}}}\,
 <{\sigma_N}{\left({\phi_j}\right)}>
  \quad     =     \quad   - \,
 <{{\sigma_{N - 1}}{\left({\phi_j}\right)}}\,
    {\phi_i}>
\ee
}
Also in this case, setting the index $i  =  0$,
we get the {\it puncture equation}:
{\boldmath
\be
{{\partial}\over{\partial{t_{0}}}}
<{\sigma_N}{\left({\phi_j}\right)}>  \,
=  \,     - \,  <{{\sigma_{N - 1}}{\left({\phi_j}\right)}}\,
   {\phi_0}>  \,   =  \,    - \,
 <{{\sigma_{N - 1}}{\left({\phi_j}\right)}}  \, {\bf P}>
\ee
}
which is the {\it quantum version}  of the
operator  equation  \, $
{{\partial}\over{\partial{t_{0}}}}\,
 {\sigma_N}{\left({\phi_j}\right)}  \quad     =  \quad   - \,
  {{\sigma_{N - 1}}{\left({\phi_j}\right)}}$
derived earlier. Further the above equation also tells us
that in the special case when all the couplings to the
 gravitational descendants vanish and the {\it puncture
operator} reduces  to the {\it identity} operator:
\bea
\label{proprr}
{{\partial}\over{\partial{t_{0}}}}\,
 {{<{\sigma_{N + 1}}{\left({\phi_i}\right)}>}_{\gf}}
  \quad  &  =  &  \quad    - \,
 <{{\sigma_{N}}{\left({\phi_i}\right)}}>\no\\
{\impl}  \quad {{\partial}\over{\partial{t_{0}}}}\,
{ {{\Bbb R}_{(N)}^i}{\left({\underline t}\right)}_{\gf}}
    \quad  &  =  &  \quad    - \,
 { {{\Bbb R}_{(N - 1)}^i}{\left({\underline t}\right)}_{\gf}}
\eea
which is exactly the same as derived
earlier from a different standpoint.

\item  \framebox[.6in]{$M = 0$} :  In
this case that above equation tells us:
\bea
{{\partial}\over{\partial{t_{N,i}}}}\,
< {\phi_j} >
  \quad   &  =   &  \quad
 <{{\sigma_{N - 1}}{\left({\phi_i}\right)}}\,
    {\phi_j}>\no\\
{\impl}    \quad {{\partial}\over{\partial{t_{1,i}}}}\,
< {\phi_j} >
  \quad   &  =   &  \quad    <{{\sigma_{0}}{\left({\phi_i}\right)}}\,
     {\phi_j}>\no\\
 &  =   &  \quad    <  {\phi_i}  \,{\phi_j}>\no\\
&  =   &  \quad  {\delta_{i + j, k}}\no
\eea
So that inverting the above equation gives:
\bea
 < {\phi_j} >  \quad  &  =   &  {\int} d{t_{1,i}}\,
  {\delta_{i + j, k}}\no\\
&  =  &   {t_{1,{k - j}}}\no
\eea
giving us the result:
{\boldmath
\be
{t_{1,i}}  \quad  =  \quad
 < {\phi_{k - i}} >
\ee
}
which gives us an explicit expression for
the couplings to the lowest order gravitational descendants.

\item  \framebox[.6in]{$N = 1$} : In this case
the eq.({\ref{opun}}) reduces to:
\be
{{\partial}\over{\partial{t_{1,i}}}}\,
 <{\sigma_M}{\left({\phi_j}\right)}>
  \,      =      - \,  <{{\sigma_{M - 1}}
{\left({\phi_j}\right)}}\,   {{\sigma_{1} }
{\left({\phi_i}\right)}}>  \,   +  \,
  <{{\sigma_{M }}{\left({\phi_j}\right)}}\, {\phi_i}>
\ee
so that setting $i = 0$,  gives us the
{\it dilaton equation}:
\be
{{\partial}\over{\partial{t_{1,0}}}}
 <{\sigma_M}{\left({\phi_j}\right)}>
  \,  \,      =    \,    - \,
 <{{\sigma_{M - 1}}{\left({\phi_j}\right)}}\,
 {\Phi_0}   >  \,   +  \,
   <{{\sigma_{M }}{\left({\phi_j}\right)}}\, {\bf P}>
\ee
where $  {{\sigma_{1} }{\left({\phi_0}\right)}}
 \, =  \, {\Phi_0}  $ is the dilaton operator and
$t_0$ is the corresponding coupling and the above
 equation shows how the dilaton and the puncture
operators get inserted into the correlation functions.

\een

\subsection{Further  Generalizations}

Now that we have obtained the equation for \,
${{\partial}\over{\partial{t_{N,i}}}}\,
 <{\sigma_M}{\left({\phi_j}\right)}>$,\,
we may  proceed towards  obtaining the similar
expressions  for the quantities like   \,
$   {{\partial}\over{\partial{t_{N,i}}}}\,
 < {\sigma_M}{\left({\phi_j}\right)}
 {\sigma_L}{\left({\phi_m}\right)}  >, \, \,
    {{\partial}\over{\partial{t_{N,i}}}}\,
  < {\sigma_M}{\left({\phi_j}\right)}
 {\sigma_L}{\left({\phi_m}\right)}
{\sigma_P}{\left({\phi_l}\right)}>, etc.$
    \,  and the higher order correlation functions.
 However,  instead of  going
 through similar computational procedures,
 we may take an alternative shortcut route as follows.

We have already seen that the quantities
 $  {{\Bbb R}_{(N)}^l}{\left({\underline t}\right)} $
      satisfy the important equation
({\ref{propr}}).  Using the result ({\ref{resa}}), we get
$$
 {{\partial^2}\over{\partial{t_i}}{\partial{t_j}}} \, \,
   {{\Bbb R}_{(N)}^l}{\left({\underline t}\right)}
 \quad    =      {{\partial}\over{\partial{t_j}}}  \,{\biggl[ \,
    <{{\sigma_{N}}{\left({\phi_l}\right)}}
 \, {\phi_i}>  \, \biggr]}
$$

Substituting the above results into the previous
equation we get the identity:

\be
\label{useful}
{{\partial}\over{\partial{t_j}}}  \,
  <{{\sigma_{N}}{\left({\phi_l}\right)}} \, {\phi_i}>  \,
   \quad   =  \quad   - \,  {\sum_{m = 0}^k}\,
{{{\cal C}_{ij}}^m} \,
 <{{\sigma_{N - 1}}{\left({\phi_l}\right)}} \, {\phi_m}>
\ee
We have already seen that the 2-point function
{\it factorizes} nicely. Using this wisdom we then have:
\bea
{{\partial}\over{\partial{t_l}}} \,
 <{{\sigma_{N}}{\left({\phi_i}\right)}} \,
 {{\sigma_{M}}{\left({\phi_j}\right)}} >  \,
  &  =  &   {\sum_{s,n}}\,  \,
  {{\partial}\over{\partial{t_l}}} \,    {\biggl[  \,
  <{{\sigma_{N}}{\left({\phi_i}\right)}} \, {\phi_s}> \,
 {\eta^{sn}_{(0)}}\,     <  {\phi_n}\,
{{\sigma_{M}}{\left({\phi_j}\right)}}> \, \biggr]}\no\\
 &  =  &   {\sum_{s,n}}\, {\eta^{sn}_{(0)}}\,
  <{{\sigma_{N}}{\left({\phi_i}\right)}} \, {\phi_s}> \,
  {{\partial}\over{\partial{t_l}}} \,
   <  {{\sigma_{M}}{\left({\phi_j}\right)}} \,
  {\phi_n}>\no
\eea
Using  ${\eta^{pm}_{(0)}}{\phi_m}  \, =  \,
{\phi_{k - p}}$, and the result from eq.({\ref{useful}}),
 we finally have the following useful {\it generalization}
 of our recursion relation to the case of the 2-point function:
{\boldmath
\bea
\label{tpun}
{{\partial}\over{\partial{t_l}}}
 <{{\sigma_{N}}{\left({\phi_i}\right)}} \,
 {{\sigma_{M}}{\left({\phi_j}\right)}} >  \,
&  =  &   - \, {\sum_{m,n}}  \,
  {{{\cal C}_{ln}}^m} \,{\biggl[}
<{{\sigma_{N}}{\left({\phi_i}\right)}} \,
 {\phi_{k - n}}>
 <{{\sigma_{M - 1}}{\left({\phi_j}\right)}} \, {\phi_m}>\no\\
& \quad &  \quad   \quad  +  \,
  <{{\sigma_{M}}{\left({\phi_j}\right)}} \,
 {\phi_{k - n}}>   <{{\sigma_{N - 1}}{\left({\phi_i}\right)}} \,
 {\phi_m}>{\biggr]}
\eea
}
We may now make the following conclusions
 from the above equation.

\ben

\item   Setting   $M =  N  =  0$, in the above
equation we get:
$${{\partial}\over{\partial{t_l}}} <{\phi_i}{\phi_j}>
  \quad  =  \quad 0
$$
which just reminds us that the 2-point function is
${\underline t}$- independent.

\item  Setting the indices  $M = 0$,  and $N = 1$
in the identity ({\ref{useful}}), we have:
\bea
 {{\partial}\over{\partial{t_l}}}  \,
 <{{\sigma_{1}}{\left({\phi_i}\right)}} \, {\phi_j}>  \,
   \quad  &   =  &  \quad   - \,  {\sum_{m = 0}^k}\,
 {{{\cal C}_{lj}}^m} \,   < {\phi_i}  \, {\phi_m}>\no\\
&  =   & \quad  -  \,  {{{\cal C}_{lj}}^{k - i}}
\eea
But   $ <{{\sigma_{1}}{\left({\phi_i}\right)}} \,
{\phi_j}>  \, =  \, {{\partial}\over{\partial{t_j}}}
 {{\Bbb R}_{(1)}^i}{\left({\underline t}\right)}$,
which when combined with the above equation gives us the result
$$
{\left( {{\partial^2}\over{{\partial{t_j}}{\partial{t_l}}}}\right)} \,
  {{\Bbb R}_{(N)}^l}{\left({\underline t}\right)}  \quad  = \quad
 - \,  {{{\cal C}_{lj}}^{k - i}}
$$
a result that we had also obtained earlier (See Corollary
Sect.{\bf 4}).

\item   Setting $l \, =  \, 0$ in  eq.({\ref{tpun}}), we get:
\bea
{\left( -  {{\partial}\over{\partial{t_0}}}\right)} \,
  <{{\sigma_{N}}{\left({\phi_i}\right)}} \,
 {{\sigma_{M}}{\left({\phi_j}\right)}} >  \, &  =
&  \, {\sum_m} \,  {\biggl[} <{{\sigma_{N}}{\left({\phi_i}\right)}}
 \, {\phi^{m}}>   <{{\sigma_{M - 1}}{\left({\phi_j}\right)}}
 \, {\phi_m}>\no\\
&  \quad &  \quad   \quad  +  \,
  <{{\sigma_{M}}{\left({\phi_j}\right)}} \, {\phi^{m}}>
  <{{\sigma_{N - 1}}{\left({\phi_i}\right)}} \,
 {\phi_m}>{\biggr]}\no\\
&  =  &    {\biggl[} <{{\sigma_{N}}{\left({\phi_i}\right)}} \,
 {{\sigma_{M - 1}}{\left({\phi_j}\right)}} >\no\\
&  \quad  &  \quad \quad  +  \, \,
 <{{\sigma_{N - 1}}{\left({\phi_i}\right)}} \,
 {{\sigma_{M}}{\left({\phi_j}\right)}} >{\biggr]}
\eea
which is a generalized form of the {\it puncture equation}.

\een

\medskip

\nd
Thus working along similar lines  ({\it i.e.}
 proceeding  with computations like those leading
to ({\ref{opun}}) and ({\ref{tpun}}) as previously
shown), one may then prove the following generalized
 operator  form of the {\it puncture equation}:

\be
\label{gpun}
{\left( - \, {{\partial}\over{\partial{t_0}}}\right)}
  \,  < {\prod_{j}} \,
 {{\sigma_{{N_j}}}{\left({\phi_{n_j}}\right)}}>
 \quad  =  \quad   {\sum_l} \,
 <   {{\sigma_{{N_l} - 1}}{\left({\phi_{n_l}}\right)}}
  \,  {\prod_{j {\not =} l}} \,
   {{\sigma_{{N_j}}}{\left({\phi_{n_j}}\right)}}>
\ee
Thus we have an alternative operator method of
 derivation of the puncture equation which has
been previously  ({\cite{{DVV1},{DVV2},{VV},{DW}}})
 derived  using an entirely  different  procedure.
 Finally, we can make an important observation here.
 In the next section we shall see that the coordinates
 in the large phase space can eventually be expressed
 in terms of those in the small phase space. Thus
having done this,  we can easily generalize the
recursion relations like (\ref {tpun}) to the entire phase space.

\bigskip


\section{{ {Relation between $t_{N,i}$  and  $t_{0,i}$ \,}}  }
\setcounter{equation}{0}

Finally, in this section we focus our attention to
the central conclusion of our present work.  We
investigate the properties of the coordinates of the
 {\it large phase space}  of our gravity-coupled model
 and we try to {\it determine} their explicit forms.
 We may recall that we  began by considering the minimal
 action perturbed by {\it all} possible fields belonging
 to the family of gravitational descendants of the primary
fields in our theory. These perturbing fields couple by
{\it a priori arbitrary} couplings $t_{N,i}$ (the
 ``coordinates'' on the {\it large phase space}).
These couplings are initially  completely unconstrained
 apart from the demand of reparametrization invariance.
 However, in the gravity-free case, it is known that
 due to certain underlying general properties of TCFT,
 these couplings can not just be anything arbitrary, but
 are highly constrained. In fact, these have to be
necessarily completely determined functions (see eqn.
 ({\ref{coupling}}))  of the {\it flat-coordinates} --
 and are hence are  no longer arbitrary at the end of
the day. In this case, the  fact that the 2-point function
 (which also defines the metric in this case) of the theory
remains unaltered even in the presence of perturbations
 leads to the explicit determination of all the couplings
 (the ``coordinates'' of the {\it small phase space}) to
 the  perturbing chiral primaries of the theory. After
 coupling this  model to gravity, we then have a much
enlarged (infinite-dimensional) phase space. In the
existing literature, not much has been said about the
properties of these  couplings to the gravitational
descendants. In particular, with gravity switched on
we have an infinite number of {\it a priori arbitrary}
  couplings (to the gravitational descendants)  in the
theory, and it is interesting and quite relevant to ask
ourselves {\it what} we can say about these couplings,
and can we {\it determine}  these couplings explicitly,
 just as we did in the gravity-free case.

Our analysis in this section will  seek to obtain the
answers for this question in the affirmative. In fact we
shall see that the answer to both these questions is {\it yes}.
The underlying topological structure of the gravity-coupled model
 {\it does},   in this case,  also lead to the complete and
explicit (modulo constant translations) determination   of
{\it all} the couplings to the gravitational descendants of
 the theory. We shall see that one can determine these
infinite set of couplings in the form of known (but
infinite in number) functions of the coordinates in
the small phase space. Further, the coordinates in the
small phase space are, in turn,  known in terms of the
flat coordinates. Thus eventually, we get to determine
all the couplings explicitly.

Let us now try to obtain this  relation between the
perturbation parameters ($t_{N,i}$) ({\ref{coup}})  of
the general gravity-coupled model  and  those  ($t_{0,i}
\, {\equiv}  \, {t_i}$)   when  the gravity is {\it switched
off}.  The couplings to the chiral primaries ({\it i.e.} our
 flat coordinates in the {\it small phase space}) are already
 known, being given by ({\ref{coupling}}). So the remaining
 task is to solve for the couplings to the gravitational
descendants. We argue that the couplings to the gravitational
 descendants {\it can}  also be determined completely using
 the informations we already have. Once we have done this,
it means that now the entire phase space of the theory is known.

We shall achieve  this by   trying to evaluate the 2-point
function
$<{\sigma_M}{\left({\phi_i}\right)}{\sigma_N}{\left({\phi_j}\right)}>$
   of the general gravity-coupled model in two different ways
and then comparing  the results to obtain the required relation.
  One way of obtaing the expression  for the above correlation
function has already been discussed in Sect. 4.3.  We now
evaluate the same in an alternative method, without resorting
to the {\it reduction  formula}  ({\ref{redfor}}) for the
gravitational descendants.   Thus following the prescription
of {\it Vafa} {\cite{VA}}  we have (the details of the
derivation ({\ref{vafa}}) are in the appendix):
\be
\label{vafa}
<{\sigma_M}{\left({\phi_i}\right)}{\sigma_N}{\left({\phi_j}\right)}>
 \, \quad \, \,  =  \quad   {{\partial}\over{\partial{t_{N,j}}}} \,
 {{\Bbb R}_{(M)}^i}{\left({\underline t}\right)}  \quad
+  \quad   {{\partial}\over{\partial{t_{N - 1,j}}}}  \,
 {{\Bbb R}_{(M + 1)}^i}{\left({\underline t}\right)}
\ee

Now writing
\bea
 {{\partial}\over{\partial{t_l}}}  \quad  \, &   =   &
 \quad  {\sum_{N,j}}\,
{\left({{\partial{t_{N,j}}}\over{\partial{t_l}}
}\right)}{{\partial}\over{\partial{t_{N,j}}}}\no\\
&   =   &  \quad  {\sum_{N,j}}\,
 {{\left({{\Bbb D}^{(N)}}\right)}_{jl}}
 {\left( {{\partial}\over{\partial{t_{N,j}}}}\right)}\no
\eea
where we have introduced  $N$ (where $N =
0,1, 2, {\ldots} {\infty}$)  matrices (each of size
$k {\times} k$)\,  ${\Bbb D}^{(N)}$ defined by:
\be
{{\Bbb D}^{(N)}_{ij}}   \quad    \stackrel{\rm def}{=}
  \quad  {\left({{\partial{t_{N,i}}}\over{\partial{t_j}}}\right)}
 \quad  {\equiv}  \quad
 {\left({{\partial{t_{N,i}}}\over{\partial{t_{0,j}}}}\right)}
  \quad \quad \quad  \quad \quad \quad (0 \, {\leq} \, i,j \, {\leq} \, k)
\ee
 so that we may write:
$$
{{\partial}\over{\partial{t_{N,j}}}}  \quad  =  \quad
   {\sum_l}\, {{\left({{\Bbb D}^{(N)}}\right)}^{-1}_{lj}}
 {\cdot}  {{\partial}\over{\partial{t_l}}}
$$
\nd
Thus, {\it determining} the matrices
 ${\Bbb D}^{(N)}$ explicitly amounts
to {\it knowing\footnote{Having done
 this, the remaining task is quite easy,
as we only need to solve these first-order differential
  equations --- to eventually express  the couplings
$t_{N,i}$s  in terms of the $t_i$s.}  the couplings}
$t_{N,i}$ in terms of $t_{i}$; and as the latter are
 already known in terms of the superpotential, this
information is sufficient to determine {\it all} the
couplings. Clearly we have:
$$
{{\Bbb D}^{(0)}_{ij}} \quad  =  \quad {\delta_{ij}}
$$
Using these matrices introduced above, we can recast
 our expression for the 2-point function as:
\bea
<{\sigma_M}{\left({\phi_i}\right)}{\sigma_N}
{\left({\phi_j}\right)}> \quad  \, \,   &   =   &
 \quad   {\sum_l}\,
 {{{\left({{\Bbb D}^{(N)}}\right)}}_{lj}^{- 1}} \,
 {{\partial}\over{\partial{t_{l}}}} \,
 {{\Bbb R}_{(M)}^i}{\left({\underline t}\right)}\no\\
&  \quad &  \quad \quad  \quad  \quad +  \quad
  {\sum_l}\,
 {{{\left({{\Bbb D}^{(N - 1)}}\right)}}_{lj}^{- 1}} \,
 {{\partial}\over{\partial{t_{l}}}} \,
 {{\Bbb R}_{(M  + 1)}^i}{\left({\underline t}\right)}
\eea

\nd
Comparing eqs.(3.27) and (3.37)  we  can
then have the following equation:
\bea
- \,   {\sum_l}\,  {\Biggl[ \, {{{{\left({{\Bbb D}^{(N)}
   }\right)}}_{lj} ^{- 1}}} \, -  \,{\left({{{\partial}
 {{\Bbb R}_{(N)}^j}}\over{\partial{t_{k - l}}}}\right)}
  \, \Biggr]}   {\left({{\partial}\over{\partial{t_{l}}}}
 \, {{\Bbb R}_{(M)}^i}{\left({\underline t}\right)}\right)}
 &  \, &        \no\\
 =  \,  & \,  &  {\sum_l}   \,   {{{{\left({{\Bbb D}^{(N - 1)}
  }\right)}}_{lj} ^{- 1}}} \,{\left( {{\partial}\over{\partial{t_{l}}}}
 \, {{\Bbb R}_{(M)}^i}{\left({\underline t}\right)}\right)}
\eea
which may be in a more convenient
notation\footnote{The summation convention
is assumed over the repeated small indices $i,l, \, etc.$} as:
\bea
{\Biggl[  \,   \,\, {\left( {{\partial}\over{\partial{t_l}}}
 \, {{\Bbb R}_{(M + 1)}^i}{\left({\underline t}\right)}\right)}
       \,{\left( {{\partial}\over{\partial{t_{k - l}}}} \,
 {{\Bbb R}_{(N - 1)}^j}{\left({\underline t}\right)}\right)}\Biggr]}
 &  =  &   - \,  \,  {{\Delta}^{(N)}_{lj}}{\left(
 {{\partial}\over{\partial{t_{l}}}} \,
 {{\Bbb R}_{(M)}^i}{\left({\underline t}\right)}\right)}   \no\\
& \quad &  \quad \quad \quad   - \,  \,
  {{\Delta}^{(N - 1)}_{lj}}{\left( {{\partial}\over{\partial{t_{l}}}} \,
 {{\Bbb R}_{(M + 1)}^i}{\left({\underline t}\right)}\right)}  \no\\
{\hbox{where}}  \quad  \quad       {{\Delta}^{(N)}_{lj}}   &
   \dfn   &     {\Biggl[ \, {{{{\left({{\Bbb D}^{(N)}
  }\right)}}_{lj} ^{- 1}}} \, -
 \,{\left({{{\partial} {{\Bbb R}_{(N)}^j}}\over{\partial{t_{k
- l}}}}\right)}  \, \Biggr]}\no\\
&  \quad  &  \quad \quad \quad ({\hbox{And clearly,}} \quad
 {{\Delta}^{(0)}_{lj}} \quad =  \quad  0.)\no
\eea
 And finally, introducing the operators  \,
 ${{\Omega}^{(N)}_{j}}$ \, defined by:
$$
{{\Omega}^{(N)}_{j}}{\left({\underline t}\right)}
\dfn  {\sum_l} \,  {{\Delta}^{(N)}_{lj}} \,
 {{\partial}\over{\partial{t_l}}}
$$
we may write the {\it equation determining
 the relation between  the couplings}:

{\boldmath
\bea
\label{rel1}
{\sum_l} \,  {\Biggl[
{\left( {{\partial}\over{\partial{t_l}}}
 {{\Bbb R}_{(M + 1)}^i}{\left({\underline t}\right)}\right)}
 {\left( {{\partial}\over{\partial{t_{k - l}}}}
 {{\Bbb R}_{(N - 1)}^j}{\left({\underline t}\right)}\right)}\Biggr]}
  \,    &  =  &   \,     -  \, \,
 {\Biggl[} \,  {{\Omega}^{(N)}_{j}}
  {{\Bbb R}_{(M)}^i}{\left({\underline t}\right)}\no\\
&   &       \quad \quad \quad    +
   \quad     {{\Omega}^{(N - 1)}_{j}}
 {{\Bbb R}_{(M + 1)}^i}{\left({\underline t}\right)}    \,  {\Biggr]}
\eea
}

\nd
Since    \, ${{\Omega}^{(0)}_{j}}  \, \, =
  \, \, 0$,   the above equation can  therefore
be used to recursively determine all the  \,
${{\Omega}^{(N)}_{j}}$s and hence finally the
 $N$ matrices  ${\Bbb D}^{(N)}$-- and hence
eventually, the desired relationship.  Even
though the above equation does not look very
 friendly at first sight, we can in fact solve
it setting specific values for $N$ and obtain
the expressions for the quantities  ${\Bbb D}^{(N)}$.
To make matters simple, we may without any loss
of generality, set $M = 0$ in the above equation
 which then reduces\footnote{where we have used the
 fact that  \,   ${{\Omega}^{(N)}_{j}}
 {{\Bbb R}_{(0)}^{k - i}}{\left({\underline t}\right)} \,
 \,  =   \, \, {{\Delta}^{(N)}_{ij}}  $}     to the following
simplified form:

\bigskip

\be
\label{rel}
{\Biggl[} \,  {{\Delta}^{(N)}_{ij}}
+   \, \,    {{\Omega}^{(N - 1)}_{j}}
 {{\Bbb R}_{(1)}^{k - i}}   \,  {\Biggr]}
 \quad  +  \quad  {\sum_l} \, {\Biggl[
  {\left( {{\partial}\over{\partial{t_l}}}
 {{\Bbb R}_{( 1)}^{k - i}}\right)}
     {\left( {{\partial}\over{\partial{t_{k - l}}}}
 {{\Bbb R}_{(N - 1)}^j}  \right)}\Biggr]}  \quad  =  \quad 0
\ee

\bigskip

\nd
with the additional constraints:  ${\Delta}^{(0)}
 \, =  \, {\Omega}^{(0)} \, = \, 0$

\bigskip

\begin{itemize}

\item  {\bf Note:}
If further we sum both sides of  eq. ({\ref{rel1}})
 over $M =  - {\infty}$ to ${\infty}$, we can write
 the above equation in terms of the quantity ${\Psi}$
defined in ({\ref{defs}}), as:

\bea
{\left[ \,  {{\Omega}^{(N)}_{j}}   \,    +   \,
 {{\Omega}^{(N - 1)}_{j}}  \,
 \right]} {{\Psi}^{(i)}}{\left({\underline t}\right)}
  &     &     \no\\
&  = &    - \, {\Biggl[
  {\left( {{\partial}\over{\partial{t_l}}}
  {{\Psi}^{(i)}}   \right)}
 {\left( {{\partial}\over{\partial{t_{k - l}}}}
 {{\Bbb R}_{(N - 1)}^j}{\left({\underline t}\right)}\right)} \Biggr]}
\eea
which we may rewrite in a more convenient form as:

\be
{\sum_l} \, {\left[ \,  {{\Delta}^{(N)}_{lj}}   \,
   +   \,    {{\Delta}^{(N - 1)}_{lj}}    \,  +
 \,   {\left( {{\partial}\over{\partial{t_{k - l}}}}
 {{\Bbb R}_{(N - 1)}^j}{\left({\underline t}\right)}\right)}
    \,   \right]}  \,
 {\left( {{\partial}\over{\partial{t_l}}}  {{\Psi}^{(i)}}
   \right)} \quad  = \quad  0
\ee
which can be interpreted as a sort of a {\it flow}
  equation in the space of couplings (the small phase space).

\end{itemize}

\bigskip

\bigskip

\nd
Thus we see that even in the presence of
gravity, one can say quite a lot about the
corresponding couplings.  The couplings to
the gravitational descendants  cannot  just
 be anything arbitrary, but are highly constrained
by the underlying topological symmetries of the theory.
 The basic symmetries of the theory are stringent
enough to completely determine the functional form
of all these couplings. Since the couplings to the
 descendant fields are now determined in terms of
those to the chiral primaries and these in turn in
terms of the {\it flat coordinates}, we see eventually
that all the informations about the large phase space
can be extracted from the superpotential characterising
the matter sector of the theory. Even though the above
 observation is based on our calculations with a specific
kind of matter model, we believe that these conclusions
should hold true quite generally ({\it i.e.}  for other
 models as well). This therefore lends credence to the
view that $2d$ topological gravity is indeed an {\it induced} effect.

\bigskip

\nd
{\bf  Some Special Cases}: \,  Let us now try to solve
 our master equation explicitly for some special cases to
show that this formalism indeed provides us with the necessary
 framework for determining  the coordinates of the large phase
space. As our formalism is of recursive nature, we need to
proceed from the  couplings to the lowest order gravitational
 descendants.

\nd
Setting $N  \, = \, 1$ in the   equation ({\ref{rel}}),
 we get

\bea
{{\Delta}^{(1)}_{ij}}   \, \, +  \, \,   {\sum_l} \,
  {\left( {{\partial}\over{\partial{t_l}}}
 {{\Bbb R}_{( 1)}^{k - i}}\right)}
    {\left( {{\partial}\over{\partial{t_{k - l}}}}
  {{\Bbb R}_{(0)}^j}  \right)}  &  =  &  0\no\\
{\impl} {{\Delta}^{(1)}_{ij}}  &  =  &  -  \,
 {\left( {{\partial}\over{\partial{t_j}}} \,
{{\Bbb R}_{( 1)}^{k - i}}{\left({\underline t}\right)}\right)}
\eea
This eventually gives us:
\bea
\label{solt1}
{{{{\left({{\Bbb D}^{(1)}   }\right)}}_{ij} ^{- 1}}}
\quad  & =  &  \quad  {\sum_l} \,  \,  {\biggl[\,
{\left({{{{\partial}} \over{\partial{t_{l}}}}} {{\Bbb R}_{(1)}^j}
   \right)}  \, -  \,  {\left( {{\partial}\over{\partial{t_j}}} \,
 {{\Bbb R}_{( 1)}^{l}}  \right)} \, \biggr]}
 {\left( {\delta_{i + l, k}} \right)}\\
&  =  &  \quad   {\sum_l} \,  \, {e_{mlj}} \,
 {\delta_{i + l, k}} \,
 { {\left( {\vek {{{\grd}_{{\underline t}}}}} \,
  {\times} \,  {{\vek  {{{\Bbb R}}}}_{(1)}} \right)}_m}
\eea
(where  ${{\vek  {{{\Bbb R}}}}_{(1)}}   \,   =
  \, {\lbrace}\,  {{\Bbb R}_{(1)}^i} \, {\rbrace}$
 is interpreted as a  vector in the
$ (k + 1)$-dimensional space of chiral primaries,
and the curl is taken with respect to the couplings
 ${t_i}$ treated as coordinates {\it i.e.} ${\vek  t} \,
= \, {\lbrace} {t_i} {\rbrace}$)  from which the desired
relation between ${t_{1,i}}$ and ${t_i}$   may be obtained.
 We may then put back the obtained values in the equation
to solve for the next order couplings.  Thus recursively
 the entire family of couplings ${\lbrace {t_{N,i}} \rbrace}$
 can be determined.

Similarly, setting $N  \, = \, 2$ in the   equation
({\ref{rel}}),  and using the previous results we can
obtain the following result:

\be
{{\Delta}^{(2)}_{ij}}  \quad  =  \quad {\sum_l} \,
{\left({{{\partial \,
 {{\Bbb R}_{(1)}^{k -i}}}}\over{\partial {t_{k - l}}}}\right)}
{\Biggl[     \, {{{\partial \,
 {{\Bbb R}_{(1)}^{l}}}}\over{\partial {t_j}}}   \,
-  \,  {{{\partial \, {{\Bbb R}_{(1)}^{j}}}}\over{\partial {t_l}}}
\, \Biggr]}
\ee
Thus introducing the antisymmetric  tensor
$V^{(N)}_{ij}$ defined by
\be
V^{(N)}_{ij}  \quad  {\dfn}  \quad  {\Biggl[
    \, {{{\partial \,
 {{\Bbb R}_{(N)}^{i}}}}\over{\partial {t_j}}}   \,
-  \,  {{{\partial \, {{\Bbb R}_{(N)}^{j}}}}\over{\partial {t_i}}}
 \, \Biggr]}
\ee
(which trivially satisfies \, ${\partial_l}V^{(N)}_{ij}
\, +  \, {cyclic \, \, \,  perm.} \, \, = \, \, 0$),
we see that the final expressions for the matrices
${{\left( {{{{\left[ \, {{\Bbb D}^{(N)}
  } \, \right]}}^{- 1}}}\right)}_{ij}}$
can be written more compactly as:

\bea
{{\left( {{{{\left[ \, {{\Bbb D}^{(1)}   } \,
 \right]}}^{- 1}}}\right)}_{ij}}      &  =   &
         {\sum_l} \,   {\left({{{\partial \,
 {{\Bbb R}_{(0)}^{k - i}}}}\over{\partial {t_{k - l}}}}\right)}
  \, V^{(1)}_{jl}  \quad  =  \quad
   {\sum_l} \, {{\delta}_{i + l, k}} \,
 V^{(1)}_{jl}\\
{{\left( {{{{\left[ \, {{\Bbb D}^{(2)}
 } \, \right]}}^{- 1}}}\right)}_{ij}}
    &  =   &       {\left({{{\partial \,
 {{\Bbb R}_{(2)}^{j}}}}\over{\partial {t_{k - i}}}}\right)}
 \,  +  \,    {\sum_l} \,
  {\left({{{\partial \, {{\Bbb R}_{(1)}^{k - i}}}}\over{\partial {t_{k -
l}}}}\right)}
 \, V^{(1)}_{jl}\\
{{\left( {{{{\left[ \, {{\Bbb D}^{(3)}
  } \, \right]}}^{- 1}}}\right)}_{ij}}
   &  =   &      {\left({{{\partial \,
 {{\Bbb R}_{(3)}^{j}}}}\over{\partial {t_{k - i}}}}\right)}
 \,         -  \, {\sum_l} {\left({{{\partial \,
 {{\Bbb R}_{(1)}^{k - i}}}}\over{\partial {t_{l}}}}\right)}
 {\left({{{\partial \,
 {{\Bbb R}_{(2)}^{j}}}}\over{\partial {t_{k - l}}}}\right)}
    +  \,    {\sum_l} \,
  {\left({{{\partial \,
 {{\Bbb R}_{(1)}^{k - i}}}}\over{\partial {t_{k - l}}}}\right)}
  \, V^{(1)}_{jl}\\
etc. \quad &  &\no
\eea
The knowledge of the matrix elements
${{\left( {{{{\left[ \, {{\Bbb D}^{(N)}   }
\, \right]}}^{- 1}}}\right)}_{ij}}$ allows us
to obtain the expressions for the couplings
$t_{N,i}$ in terms of the coordinates $t_{i}$
of the small phase space. The  latter are in turn
obtainable in terms of the superpotential $W$
characterizing the topological matter sector
(the $A_{k + 1}$ superpotential in our case) of
the theory. Thus though we start off with an infinite
dimensional coupling constant space, eventually the
entire phase space of couplings can be expressed as
known polynomial functions of the flat-coordinates
 based on the information about the superpotential of
the matter sector {\it only} --- {\it i.e.} we are thus able,
 at the end of the day, to  explicitly determine the
functional forms  \, ${t_{N,i}} \, {\left( {t_0}, {t_1},
{t_2}, {\ldots} , {t_k}\right)}$ \,  of  the  gravitational
 descendant couplings.


\section{Discussions and Further Outlook}

In our formalism, we   have effectively considered
 the topological minimal model $A_{k + 1}$ in a
{\it background} of gravitational descendant fields.
These descendant fields have relative strengths
determined by the couplings ${t_{N,i}}$. One can
thus  interpret this enlarged phase space of the
theory as the space of all perturbations of our
original model, with the couplings playing the role
of coordinates  in the  space of perturbations.
Further, the constraint of {\it flatness}  on the
 {\it coordinates} of the small phase space uniquely
 fixes up the couplings to the chiral primaries.
We have also seen, how by computing the 2-point
correlation functions in two different ways, namely,
  once working on the large phase space, and then
secondly   using the reduction formula for the
gravitational descendants (and hence effectively
 working on the small phase space, renormalized
by the gravity-couplings), and  finally comparing
the two results, one can also relate the coordinates
on the large phase space to those on the  small phase
space.  In effect,  this tells us that we have
succeded in reducing  the infinite number of
 {\it  a priori arbitrary}   couplings  ${t_{N,i}}$,
  to an  infinite number of {\it known}  functions
 of a {\it finite}  number of variables  ${t_{i}}$.

Using the properties of the underlying multiplicative
ring structure of the $N = 2$ theory of our matter sector,
 we have obtained a set   of orthogonal polynomials.
(These are orthogonal with respect to the inner product
on the ring). To lowest order, these {\it coincide}
with the basis of the perturbed chiral ring, but in
 higher orders  this convenient property is destroyed
as each of these polynomials  receive  contributions
 from the   polynomials of a different order, coupled
by the  coordinates  on the  large phase space. Thus
in effect these polynomials  get renormalized at each
 order, and by imposing the condition of orthogonality
at each order, one can then get an interesting hierarchy
 of differential equations, governing the
coordinate-dependences of these polynomials.
In obtaining such relations,  we have found an
 interesting operator equivalence  ({\ref{opeq}}),
 which allows us to  directly write down such equations
  at any order. Further, this operator equivalence may
 have deeper mathematical interpretaions in  along
 the lines of Saito's work {\cite{{LOS},{SAI}}}.

The  operator versions of the relations determining
 the {\it gravitaional dressing} of the correlation
functions that we have obtained in Sect.5, are
essentialy generalized Ward identities in the language
of path-integrals, and valid throughout  the large phase
 space.   We have,  in our work, considered only the two
 lowest order functions, namely the 1- and 2-point
correlation functions. Higher order correlation
functions can eventually be written down in terms
of these (using the factorization property, and the
recursion relations), and hence such generalized Ward
identities  on the large phase space,  for any N-point
 correlation function can now be constructed, thus
leading to an alternative derivation of  the Virasoro
recursion relations of topological gravity. Restricting
these identities to the small phase space gives us the
generalized {\it puncture} and {\it dilaton} equations
which have been derived previously from an entirely
different standpoint {\cite{{DVV1},{DVV2},{VV},{DW}}}.

The generalized puncture and dilaton equations are known
 to give rise to an interesting non-commutative contact
 term algebra (isomorphic to the Virasoro algebra)
 {\cite{{DVV1},{DVV2},{VV}}}. It would be  interesting
 to investigate how these commutation relations can be
obtained in our formailsm. In our formalism, we have
viewed  the model of topological gravity coupled to
topological matter as the theory of a topological
minimal matter in the presence of an infinite-dimensional
 background of gravitational fields. One can then compute
 the $\beta$ functions for the couplings to these background
 fields, and thus investigate the  {\it  multi-critical}
 behaviour of the theory. One could also, alternatively,
 use our expression for the free-energy to examine the aspects
of  {\it gravitational phase transitions} in the theory.
Because of the close relationship between the $A_{2k + 1}$
and $D_{k + 2}$ minimal models,  our above formalism can
be readily modified  to discuss coupling the   $D_{k + 2}$
  model to  $2d$ topological gravity.  Further we may easily
 consider the $k \,  {\rightarrow}  \, 0$ limit of  our
results to derive the results for {\it pure} gravity.
This limit would thus lead to a complete  solution of
pure topological gravity.


\appendix

\section{Appendix}
\setcounter{equation}{0}

\subsection{Proof of the Property (4.12) }

 Using eq. (2.38) from the previous section,
\bea
\label{appc}
b(N,l) \, {\oint}dx \, {L^{(N - 1)(k + 2) + l + 1}}
\, {\left({{\partial{\phi_i}}\over{\partial{t_j}}}\right)}
 &  =  &   - \, b(N,l) \, {\oint}dx \, {L^{(N - 1)(k + 2)
+ l + 1}} \, {{\partial_x} {{\left[ \,
{{{\phi_i}{\phi_j}}\over{W^\prime}}\, \right]}_+}}\no\\
 &  =  &    b(N - 1 ,l) \, {\oint}dx \,
 {\left({L^{(N - 1)(k + 2) + l}}{\partial_x}L\right)} \,
 { {{\left[ \, {{{\phi_i}{\phi_j}}\over{W^\prime}}\,
 \right]}_+}}\no\\
 &  =  &    b(N - 1 ,l) \, {\oint}dx   { {{\left[ \,
 {{ {\sum}  {{{\cal C}_{ij}}^m}  {\phi_m}
   }\over{W^\prime}} \right]}_+}}
  {\left({L^{(N - 1)(k + 2) + l}}{\partial_x}L\right)}\no\\
 &  =  &   {\sum_m} \, {{{\cal C}_{ij}}^m}
 { {{\left(  {{{\phi_m}}\over{W^\prime}} \right)}_+}}
 b(N - 1 ,l) \,{\oint}dx
 {\left({L^{(N - 1)(k + 2) + l}}{\partial_x}L\right)}\no\\
&  =  &  0
\eea
\samepage
where we have used the fact that \,
${ {{\left[ \, {{{\phi_m}}\over{W^\prime}} \right]}_+}}$\,
  is actually independent of $x$ (from eq.
 ({\ref{propa2}}) proved earlier)  to pull it
out of the integration.

\bigskip

\subsection{Computation of \,  $
<{\sigma_M}{\left({\phi_i}\right)}{\sigma_N}{\left({\phi_j}\right)}>$}

Following the prescription {\cite{VA}}
 of {\it Vafa}, the 2-point function is given by:
\bea
<{\sigma_M}{\left({\phi_i}\right)}
{\sigma_N}{\left({\phi_j}\right)}> \,
 \,  & \,  =   \,  &   {\oint}dx \,
{\left({{{\sigma_M}{\left({\phi_i}\right)}
{\sigma_N}{\left({\phi_j}\right)}}\over{W^\prime}}\right)}\no\\
&  \,  =  \,  &  b(M ,i) \,
  {\oint}dx \,{{{\left[{L^{M(k + 2)  + i}} \,
 {\partial_x}L\right]}_+}\over{{L^{k + 1}}{\partial_x}L}}
 {\left(  - {L^{k + 1}}
 {{\partial L}\over{\partial{t_{N,j}}}}\right)}\no\\
&  \,  =  \,  &  b(M ,i) \,
  {\oint}dx \,{{{\left[{L^{M(k + 2)
 + i}} \, {\partial_x}L\right]}}\over{{L^{k + 1}}{\partial_x}L}}
  {\left(  - {L^{k + 1}}
 {{\partial L}\over{\partial{t_{N,j}}}}\right)}\no\\
&  \,  \quad   \,  &   \quad
 \quad - \,  b(M ,i) \,
  {\oint}dx \,{{{\left[{L^{M(k + 2)  + i}}
{\partial_x}L\right]}_-}\over{{L^{k + 1}}{\partial_x}L}}
 {\left(  - {L^{k + 1}}
{{\partial L}\over{\partial{t_{N,j}}}}\right)}\no\\
&  \, =  \,  &   {{\partial}\over{\partial{t_{N,j}}}} \,
 {{\Bbb R}_{(M)}^i}{\left({\underline t}\right)}
 \quad  +  \quad \,  {\cal T}
\eea
where we have called the contribution from the
 second term ${\cal T}$, and  this is given by:
\bea
{\cal T} \quad  &   {\dfn}    &    \, \,  \,
  b(M ,i) \,   {\oint}dx \,{{{\left[{L^{M(k + 2)
 + i}}\, {\partial_x}L\right]}_-}\over{{L^{k + 1}}{\partial_x}L}}
  {\left(   {L^{k + 1}} {{\partial L}\over{\partial{t_{N,j}}}}
 \right)}\no\\
&   =    &    -  \,    b(M,i) \,
 {\oint}dx \,{{{\left[{L^{M(k + 2)  + i}}\,
 {\partial_x}L\right]}_-}    }
  {\left(   {{{\sigma_N}{\left({\phi_j}\right)}}\over{W^\prime}}
   \right)}\no\\
&   =    &    -   \,    b(M ,i) \,
   {\oint}dx \,{{{\left[{L^{M(k + 2)  + i}}\,
 {\partial_x}L\right]}_-}    } {\left[ \,
 {\int^x} {\sigma_{N - 1}}{\left({\phi_j}\right)}  \, +
 \, {\sum_m}  {\left({{\phi_{k - m}}\over{W^\prime}}\right)}
 {{\partial}\over{\partial{t_m}}}
 {{\Bbb R}_{(N)}^j}{\left({\underline t}\right)}     \right]}\no\\
&   =    &    -   \,    b(M ,i) \,
  {\oint}dx \,{{{\left[{L^{M(k + 2)
+ i}}\, {\partial_x}L\right]}_-}    } {\left[ \,
 {\int^x} {\sigma_{N - 1}}{\left({\phi_j}\right)}
 \,      \right]}\no\\
& \quad &  \quad \quad \quad ({\hbox{since we have the
 $x$- behaviour}}  \quad  {\left({{\phi_{k - m}}\over{W^\prime}}\right)}
 \, \, {\sim}  \, \, {1\over{x}}
+  \ld)\no\\
&   =    &    -   \,    b(M + 1 ,i) \,
   {\oint}dx \,\, {\partial_x}{{{\left[{L^{M(k + 2)
  + i + 1}}\, \right]}_-}    } {\left[ \,
 {\int^x} {\sigma_{N - 1}}{\left({\phi_j}\right)}  \,
     \right]}\no\\
&   =    &    \, \,    \,    b(M + 1 ,i) \,
   {\oint}dx \, {{{\left[{L^{M(k + 2)  + i + 1}}\,
 \right]}_-}    } {\left[ \,
  {\sigma_{N - 1}}{\left({\phi_j}\right)}  \,
     \right]} \quad  \quad  ({\hbox{by partial integration}}) \no\\
&   =    &    -   \,    \,    b(M + 1 ,i) \,
  {\oint}dx \, {{{\left[{L^{M(k + 2)  + i + 1}}\,
 \right]}}    } {\left( \,     {L^{k + 1}}
 {{\partial L}\over{\partial{t_{N - 1,j}}}}
       \right)}\no\\
&   =    & \,  {{\partial}\over{\partial{t_{N - 1,j}}}}
  \, {{\Bbb R}_{(M + 1)}^i}{\left({\underline t}\right)}
\eea
Hence adding up the contributions from both the terms,
 we get the desired relation ({\ref{vafa}}) used in our
earlier section.


\end{document}